\documentclass[a4paper,superscriptaddress,twocolumn,showkeys,prd, aps]{revtex4-1}

\usepackage{latexsym,amsmath}
\usepackage{amsbsy}
\usepackage[pdftex]{graphicx}
\usepackage{amssymb}
\usepackage{epstopdf}
\usepackage[svgnames,dvipsnames,x11names]{xcolor}
\usepackage[colorlinks=true,allcolors=blue]{hyperref}

\newcommand{\ZZZ}[2]{}

\begin{document}
	\title{Measuring spatial distances in causal sets via causal overlaps}
	\date{\today}
	\author{Mari\'an Bogu\~n\'a}
	\email[]{marian.boguna@ub.edu}
	\affiliation{Departament de F\'isica de la Mat\`eria Condensada, Universitat de Barcelona, Mart\'i i Franqu\`es 1, E-08028 Barcelona, Spain}
	\affiliation{Universitat de Barcelona Institute of Complex Systems (UBICS), Universitat de Barcelona, Barcelona, Spain}
	
	\author{Dmitri Krioukov}
	\affiliation{Department of Physics, Northeastern University, Boston, Massachusetts 02115, USA}
    \affiliation{Network Science Institute, Northeastern University, Boston, Massachusetts 02115, USA}
    \affiliation{Department of Mathematics, Northeastern University, Boston, Massachusetts 02115, USA}
    \affiliation{Department of Electrical and Computer Engineering, Northeastern University, Boston, Massachusetts 02115, USA}
			
\begin{abstract}
Causal set theory is perhaps the most minimalistic approach to quantum gravity, in the sense that it makes next to zero assumptions about the structure of spacetime below the Planck scale. Yet even with this minimalism, the continuum limit is still a major challenge in causal sets. One aspect of this challenge is the measurement of distances in causal sets. While the definition and estimation of time-like distances is relatively straightforward, dealing with space-like distances is much more problematic. Here we introduce an approach to measure distances between space-like separated events based on their causal overlap. We show that the distance estimation errors in this approach vanish in the continuum limit even for smallest distances of the order of the Planck length. These results are expected to inform the causal set geometrogenesis in general, and in particular the development of evolving causal set models in which space emerges from causal dynamics.

\end{abstract}
\maketitle


\section{Introduction}
\label{sec:intro}

Planck units, introduced by Planck-self~\cite{planck1899}, mark a pivotal cornerstone in the quest for natural units that rely solely on the fundamental constants rather than on experimental artifacts. At its inception, the significance of the Planck scale was not fully realized even by Planck. This scale, however, has emerged as a critical threshold, embodying the limitations of modern physics in attempting to reconcile quantum mechanics and general relativity. This problematic reconciliation revealed a fundamental incompatibility between quantum mechanics and general relativity at the Planck scale, underscored by the Heisenberg uncertainty principle~\cite{heisenberg1925} and the Schwarzschild black hole radius~\cite{schwarzschild1916}. This incompatibility asks for the formulation of a theory of quantum gravity, a pursuit that remains a fundamental challenge in physics for more than a century.

Causal sets theory (CST) stands as a promising avenue in this pursuit, proposing a conceptual framework in which spacetime at the Planck scale is discrete, consisting of fundamental spacetime atoms interconnected by causal relations~\cite{Bombelli:1987im,bombelli1988,meyer1989,sorkin1991,sorkin1997,rideout1999,sorkin2005causal,sorkin2005,henson2006,dowker2006,surya2008,henson2010,surya2012,Surya:2019aa}. This approach is motivated by the work of Hawking, King, McCarthy~\cite{hawking_king_mccarthy_1976} and Malament~\cite{malament1977}, which shows that spacetimes sharing identical causal structures are essentially equivalent up to a conformal factor.

The discretization of spacetime suggested by CST implies that our everyday experience of spacetime as a smooth continuous manifold is an illusion induced by coarse-graining at scales that are many orders of magnitude larger than the Planck scale. This transition to larger scales is known as the continuum limit of causal sets. Taking this continuum limit demands the ability to measure distances solely from the causal structure. This task is relatively straightforward for time-like separated events~\cite{bombelli1988}, but it is a significantly more intricate endeavor for space-like separated events~\cite{Brightwell:1991aa,Rideout_2009,Eichhorn_2019}.

Here, we introduce a methodology for measuring space-like distances in Minkowski spacetimes and causal sets built on them. We measure distances between space-like separated events based on causal overlaps between them, and show that such measurements remain extremely precise all the way down to the Planck scale. Furthermore, this approach provides an easy-to-work-with framework for defining reference frames and evaluating some kinematic quantities along time-like paths.

We proceed by recalling some basic background material concerning the evaluation of proper times between time-like separated events and other relevant matters in Section~\ref{sec:first}. In Section~\ref{sec:spacedistance}, we introduce our approach to measure proper distances between space-like separated events, and discuss the associated distance estimation errors, which all go to zero in the continuum limit. In Section~\ref{sec:simulations}, we perform massive numerical experiments to test our method both on long-range distances as well as on short-range ones, which are of the order of the Planck length. In Section~\ref{sec:kinematics}, we show how our results enable the definition of inertial reference frames, and how to measure some kinematic properties of time-like curves. We conclude with relevant remarks, including those concerning the possibility of extending our approach to curved spacetimes, in Section~\ref{sec:conclusions}.

\section{Connecting discrete and continuous worlds: causal sets versus Lorentzian manifolds}
\label{sec:first}

A causal set ($\mathcal{C}, \prec$) is a locally finite set with a transitive order relation defined among some of its elements, see~\cite{Surya:2019aa} and references therein. By locally finite, we mean that given two elements $x \prec y \in \mathcal{C}$, the number of elements in the set $\{z \in \mathcal{C} | x \prec z \prec y\}$ is finite. With this order relation, the causet $\mathcal{C}$ can be fully encoded as a directed acyclic graph $\mathbb{G}_{\mathcal{C}}$ defined as follows:
\begin{enumerate}
\item
    Each element $x \in \mathcal{C}$ is a node in the graph $\mathbb{G}_{\mathcal{C}}$.
\item
    There is a directed link in $\mathbb{G}_{\mathcal{C}}$ pointing from $x$ to $y$ if and only if $x \prec y$ and $\nexists z \in \mathcal{C} | x \prec z \prec y$.
\end{enumerate}
In other words, a link from $x$ to $y$ means that there is no alternative directed path in $\mathcal{C}$ from $x$ to $y$. Notice that any causal relation between two elements $a$ and $b$ in $\mathcal{C}$ (not necessarily connected in $\mathbb{G}_{\mathcal{C}}$) can be inferred from the existence or absence of a path (or chain) in $\mathbb{G}_{\mathcal{C}}$ connecting both elements. Thus, hereafter, we use the causet $\mathcal{C}$ or its graph representation $\mathbb{G}_{\mathcal{C}}$ interchangeably.

As defined above, causal sets are general mathematical structures unrelated to any physical system. However, they are particularly suitable for describing the causal structure of spacetime. Within this context, we say that for a pair of elements $x, y \in \mathcal{C}$, $x \prec y$ if and only if $y$ is in the future of $x$, or $x$ is in the past of $y$.

One connection between discrete causal sets and continuous spacetimes goes as follows. The causal structure of a Lorentzian manifold $\mathcal{M}$ defines a transitive partial order between spacetime points in the manifold. Therefore, a Poisson point process of intensity $\rho$ on $\mathcal{M}$ defines a Lorentz-invariant causal set $\mathcal{C}$. When the intensity $\rho$ diverges, one should be able to recover the original continuum manifold $\mathcal{M}$ from this causal set $\mathcal{C}$ alone. Understanding in what precise sense the continuum limits of such discrete causal sets are smooth Lorentzian manifolds is one of the major challenges within the causal set program~\cite{major2007recovering,brightwell20082d,Major_2009,Benincasa:2010eu,surya2012evidence,saravani2014causal,belenchia2016continuum,machet2020continuum}.

\subsection{Proper times in Minkowski spacetimes}

Important steps in this direction have been taken to recover the proper time between time-like separated events.

In general relativity, the proper time of an observer is defined as the time in the co-moving reference frame, where the space coordinates of the observer are fixed. In the case of a causal set, the minimum possible step for any observer is a link in $\mathbb{G}_{\mathcal{C}}$. Thus, we can assume that such links define the fundamental unit of proper time, ultimately related to the Planck time $t_P \equiv \sqrt{\frac{G \hbar}{c^5}}$. The proper time elapsed along any chain of links in $\mathbb{G}_{\mathcal{C}}$ is then proportional to the number of steps along the chain. Using these ideas, in~\cite{Brightwell:1991aa}, the time interval between two time-like separated events $a \prec b$ is defined as the number of links in the longest chain connecting $a$ and $b$, denoted as $n_{\mathcal{C}}(a,b)$, which defines the geodesic in $\mathcal{C}$ between the two events.

For causal sets sprinkled via Poisson point processes onto Minkowski ${\mathbb M}^{d+1}$ or conformally flat spacetimes of any dimension $d+1$, it has been shown that the proper time $\tau_{\cal C}(a,b)$ between any two events $a \prec b$ measured in the causal set, converges in probability to the time-like distance $\tau_{{\mathbb M}^{d+1}}(a,b)$ between the events in the spacetime,
\begin{equation}
\tau_{\cal C}(a,b)\equiv \alpha_d \rho^{-1/(d+1)} n_{\cal C}(a,b) \longrightarrow \tau_{{\mathbb M}^{d+1}}(a,b),  
\label{eq:1}
\end{equation}
as $\rho \rightarrow \infty$~\cite{Bollobas:1991aa,Brightwell:1991aa,Bachmat:2008aa}.
Here, $\alpha_d$ is a constant that depends only on the dimension of the spacetime. It is exactly known only for $d=1$, $\alpha_1=1/\sqrt{2}$, whereas for higher dimensions numerical simulations give $\alpha_1 \lesssim \alpha_2 \lesssim \alpha_3$~\footnote{Our numerical simulations seems to suggest that in the limit $\rho \rightarrow \infty$ $\alpha_1=\alpha_2=\alpha_3$.}.

Setting $n_{\cal C}(a,b)=1$ in Eq.~\eqref{eq:1} defines the characteristic unit of proper time as a function of the density of the Poisson point process as $\tau_0 \sim \rho^{-\frac{1}{d+1}}$.
Equation~\eqref{eq:1} provides a way to define an estimator of the proper time in the manifold using only the information from the causal set. More importantly, it can be shown that for large density
\begin{equation}
\tau_{\cal C}(a,b)=\tau_{{\mathbb M}^{d+1}}(a,b)+ \rho^{\frac{\beta_d-1}{d+1}} \zeta_{d}
\label{error_tau}
\end{equation}
with $\beta_d<1$ and $\zeta_{d}$ a random variable with bounded fluctuations~\cite{Bachmat:2008aa}. The exact values of the exponents $\beta_d$ are only known for $d=1$, $\beta_1=1/3$, whereas for higher dimensions take the approximate values $\beta_2 \approx 1/4$, $\beta_3 \approx 1/6$ and $\beta_d \approx 0$ for $d \ge 4$. Yet, even though the exact value of $\beta_d$ is unknown, by knowing that it is smaller than one, we observe that by just counting links in $\mathbb{G}_{\mathcal{C}}$ in the continuum limit $\rho \rightarrow \infty$, we recover the actual proper times, up to the conformal factor $\alpha_d \rho^{-1/(d+1)}$.

\subsection{Organization of links in Minkowski causal sets}

\begin{figure}
\centerline{\includegraphics[width=\columnwidth]{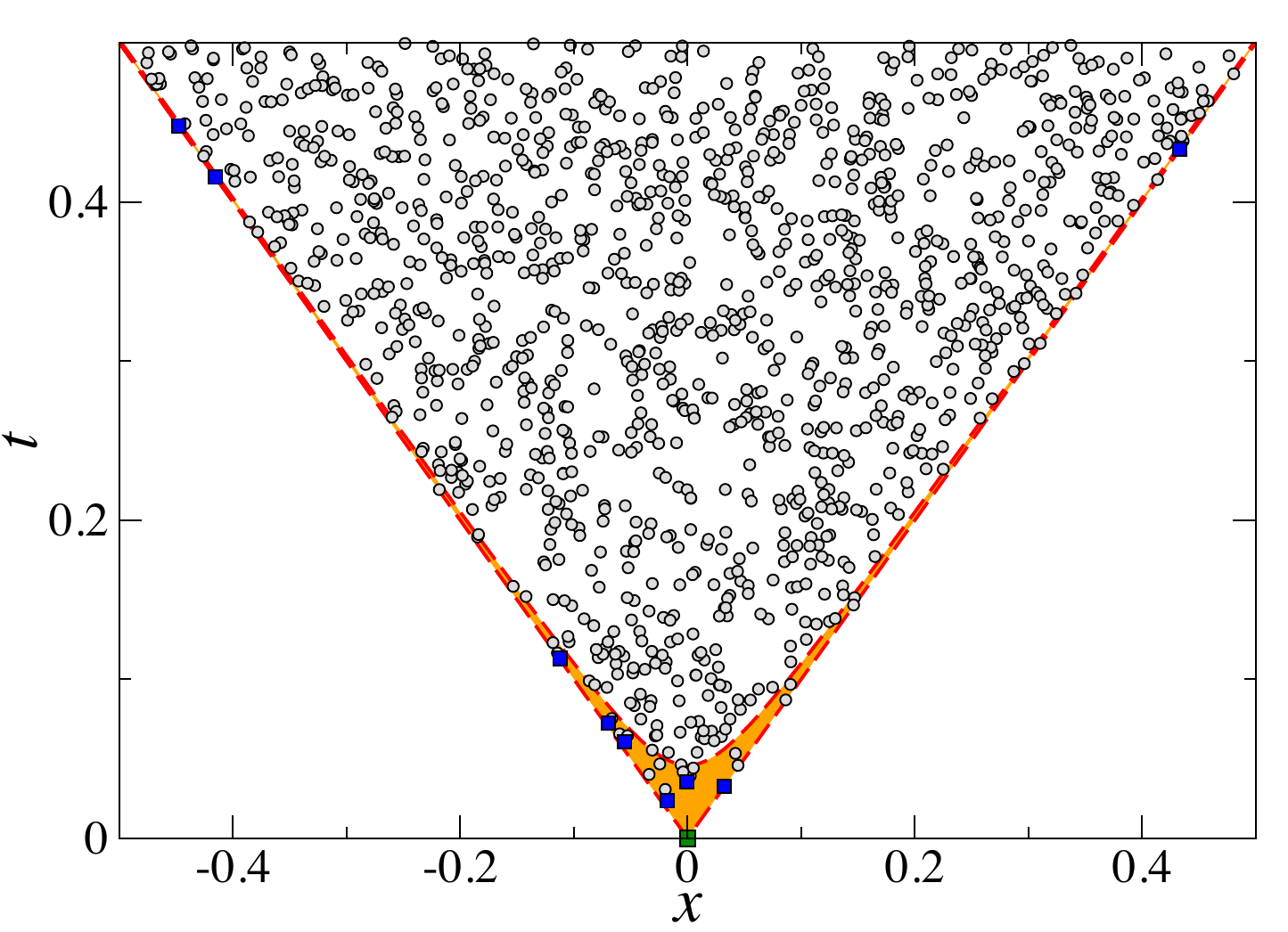}}
\caption{Direct neighbors of the event at the origin of coordinates (blue squares) in a causal set generated from patch of $\mathbb{M}_{2}$ at density $\rho=5000$. The red dashed lines are the hyperboloids of constant proper times with $\tau^{\pm}$ from Eq.~\eqref{eq:tauplusminus}. Gray symbols are events in the future of the root event but not directly connected to it.}
\label{fig:firstneighbors}
\end{figure}

Before proceeding to deriving similar results for space-like distances, we need to recall how the infinite number of links emanating from a given event in $\mathbb{G}_{\cal C}$ are distributed in $\mathbb{M}_{d+1}$~\cite{Surya:2019aa}.

Without loss of generality, let us focus on an event located at the origin of coordinates $x^i=t=0$. Its first neighbors are obviously within its future light cone and must be space-like separated, as otherwise they could not be first neighbors. In $d+1$ dimensions, the future light cone of a given point can be parametrized such that the metric tensor within the light cone can be written as
\begin{equation}
ds^2=-d \tau^2+\tau^2(d\chi^2+\sinh^{2}{\chi} d \Omega_{d-1}^2)
\end{equation}
where $d \Omega_{d-1}^2$ is the metric tensor of a $(d-1)-$dimensional sphere of unit radius, $\tau$ is the proper time, and the term within the parenthesis is the metric of the $d$-dimensional hyperbolic space of constant curvature $K=-1$. A Poisson point process with density $\rho$ means that the local density of sampled events is proportional to the volume element. Thus, the expected number of events in $\cal C$ in an infinitesimal neighborhood of coordinates $(\tau,\chi,\Omega_{d-1})$ is then
\begin{equation}
\rho dV=\rho [\tau^d d\tau][\sinh^{d-1}{\chi} d\chi d\Omega_{d-1}].
\end{equation}
However, not all events in the future light cone are direct neighbors of the root event. To calculate the density of first neighbors, we must first evaluate the probability that an event at proper time $\tau$ from the root event is actually connected to it. Given that Poisson point processes are Lorentz invariant, this probability is only a function of $\tau$ and can be computed as the probability that there is not any event within the Alexandrov interval between the root event and the event on the hyperboloid of constant proper time $\tau$. This probability reads
\begin{equation}
\mbox{Prob}(\tau)=e^{-\rho V_d(\tau)},
\end{equation}
where $V_d(\tau)$ is the volume of the Alexandrov set given by
\begin{equation}
V_d(\tau)=v_d \tau^{d+1} \;; v_d\equiv\frac{1}{(d+1) \Gamma \left( 1+\frac{d}{2}\right)} \left( \frac{\pi}{4}\right)^{\frac{d}{2}}.
\label{alexandrov}
\end{equation}
The expected number of irreducible links of the root event in an infinitesimal neighborhood of coordinates $(\tau,\chi,\Omega_{d-1})$ is then
\begin{equation}
\left[\rho \tau^d e^{-\rho V_d(\tau)} d\tau \right] \left[\sinh^{d-1}{\chi} d\chi d\Omega_{d-1}\right].
\label{eq:volumeelement}
\end{equation}
This result tells us that while first neighbors are homogeneously distributed on the hyperbolic space at constant density $[(d+1) v_d]^{-1}$, their proper time coordinates are not homogeneously distributed. Instead, they follow the normalized measure
\begin{equation}
p(\hat{\tau})=(d+1) \hat{\tau}^{d} e^{-\hat{\tau}^{d+1}},
\label{p(tau)}
\end{equation}
where we have defined the dimensionless proper time
\begin{equation}
\hat{\tau} \equiv  (\rho v_d)^{\frac{1}{d+1}}\tau
\label{eq:ptau}
\end{equation}
This equation confirms that the characteristic proper time of an individual link in $\mathbb{G}_{\cal C}$ scales with the density as $\rho^{-1/(d+1)}$. Besides, if we interpret Eq.~\eqref{p(tau)} as a probability density, we can state that, with a confidence level of $99\%$, direct neighbors of the root event are homogeneously distributed within the hyperbolic shell enclosed by the two hyperboloids of proper times 
\begin{equation}
\tau^-=\left[ -\frac{\ln{(1-p)}}{\rho v_d}\right]^{\frac{1}{d+1}} \;\mbox{and}\; \tau^+=\left[ -\frac{\ln{p}}{\rho v_d}\right]^{\frac{1}{d+1}}
\label{eq:tauplusminus}
\end{equation}
with $p=0.005$. Figure~\ref{fig:firstneighbors} shows a case example of a random sprinkling in a square patch of $\mathbb{M}_2$ at density $\rho=5000$. Highlighted in red squares the figure shows the events directly connected to the root event which, as predicted, are within the two hyperboloids of proper times $\tau^\pm$ (highlighted in orange). It is also worth mentioning that despite the fact that neighbors are homogeneously distributed in this hyperbolic shell, they are strongly correlated by the condition of being space-like separated. Besides, within this shell, neighbors are distributed at a constant density that does not depend on the sprinkling density $\rho$.

The particular form of Eqs.~\eqref{p(tau)} and~\eqref{eq:ptau} implies that both the average and standard deviation of proper times of direct neighbors of the root event scales as $\rho^{-1/(d+1)}$ so that fluctuations of the proper time of single links do not vanish in the limit $t_P \rightarrow 0$ (or  $\rho \rightarrow \infty$). This, however, does not imply that the continuum limit cannot be achieved. To see this, imagine that we take a chain of $n$ links from $\mathbb{G}_{\cal C}$ without any particular selection rule of links. The total proper time along the chain is
\begin{equation}
\tau(n)=\sum_{i=1}^n \tau_i
\end{equation}
where $\tau_i$ is the proper time of a single link of the chain. If there is no bias in selecting links, we can assume that $\tau_i$s are identical and independent random variables. Therefore, the average and standard deviation of $\tau(n)$ are given by
\begin{equation}
\langle \tau(n) \rangle=n\langle \tau \rangle \mbox{ ; } \sigma_{\tau(n)}=\sqrt{n} \sigma_{\tau},
\end{equation}
so that the coefficient of variation is
\begin{equation}
CV_{\tau(n)}=\frac{\sigma_{\tau}}{\sqrt{n} \langle \tau \rangle}
\end{equation}
If further assume that $\langle \tau(n) \rangle$ is a macroscopic (but constant) proper time in the Minkowski spacetime, then we can write that
\begin{equation}
CV_{\tau(n)}=\frac{\sigma_{\tau}}{\sqrt{\langle \tau \rangle \langle \tau(n) \rangle}}\sim \rho^{-\frac{1}{2(d+1)}} \rightarrow 0.
\end{equation}
So that in the continuum limit ($\rho \rightarrow \infty$) the graph definition of proper time along any time-like curve is exactly the same as in the manifold.

\section{Measuring proper lengths between space-like separated events using causal overlaps}
\label{sec:spacedistance}
\begin{figure}
\centerline{\includegraphics[width=\columnwidth]{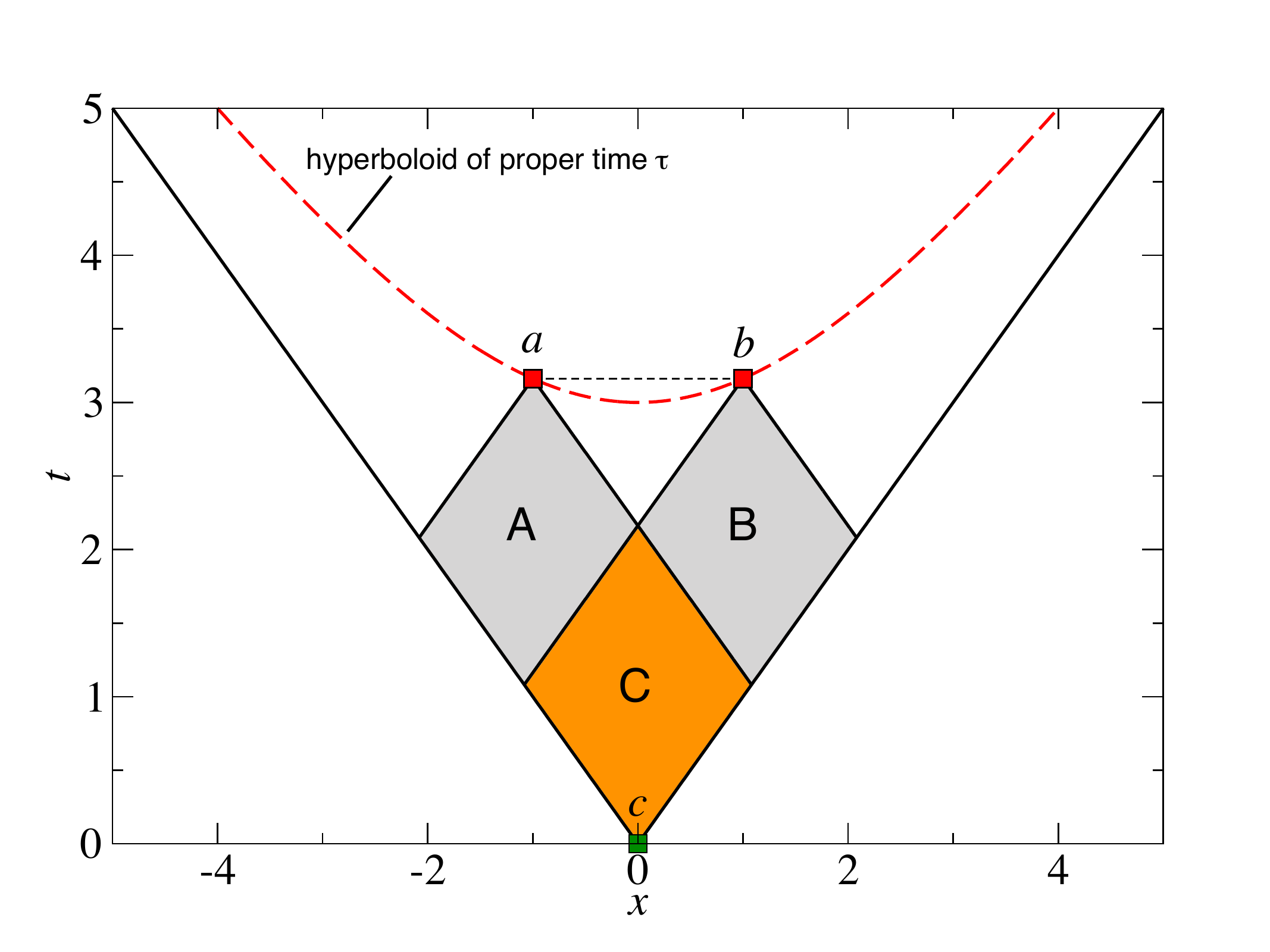}}
\caption{{\bf Definition of causal overlaps.} Sketch of the regions used for the computation of the causal overlap between events $a$ and $b$ with respect to event $c$.  Region $A$ ($B$) is the part of the Alexandrov interval between events $a$ ($b$) and $c$ that does not contain the past of event $b$ ($a$). Region $C$ is defined as the intersection of the common past of events $a$ and $b$ with the future of event $c$. The causal overlap is the ratio between the volume of region $C$ and the smallest of the volumes of regions $A \cup C$ and $B \cup C$.}
\label{fig:overlap}
\end{figure}
The evaluation of proper lengths among space-like separated events from the causal structure is far more complex than in the case of time-like separated events. This is due to the fact that since space-like events are causally unrelated, distances between them can only be defined based on the intersection of light cones. While this is not a problem in the continuum, not all possible prescriptions to evaluate geometric distances can be easily translated to discrete causets~\cite{Brightwell:1991aa,Rideout_2009}.

One of the most recent and promising approaches~\cite{Eichhorn_2019} evaluates the distance between two events $a$ and $b$ in a Cauchy hypersurface $\Sigma$ as 
\begin{equation}
d_{\Sigma}(a,b) \propto \inf_{r \in \mathcal{H}(a,b)}[V(r)]^{\frac{1}{d+1}},
\label{eq:distance1}
\end{equation}
where $\mathcal{H}(a,b)$ is the set of future events that are simultaneously null to both $a$ and $b$ and $V(r)$ the volume of the past light cone of one such point bounded from below by $\Sigma$. Notice that Eq.~\eqref{eq:distance1} is the distance calculated with the metric induced by $\mathbb{M}^{d+1}$ on $\Sigma$. While this has perfect sense in the continuum, its implementation to causal sets has some caveats. The first one concerns the fact that $d_{\Sigma}(a,b)$ depends on the choice of the Cauchy hypersurface, which in the causal set corresponds to an unextendible antichain. However, discrete Cauchy surfaces in the causal set form an uncountable infinite set and, thus, it is not clear what would be the correct choice without the help of a pre-existing embedding into a continuous manifold. The second problem is due to the absence in causal sets of events in the null surface $\mathcal{H}(a,b)$, which induces a strong error at distances of the order of the Planck scale. 

Here we introduce a distance estimator for causal sets that is able to overcome these problems. We begin not with causal sets, but with a measure of the distance between two spacelike-separated events in any spacetime based on their causal overlap. Specifically, we define the causal overlap $\mathcal{O}(a,b)$ between two events $a$ and $b$ with respect to an arbitrary event $c \in \mbox{Past}(a,b)$ in their common past as 
\begin{equation}
\mathcal{O}(a,b) \equiv \frac{V\left[C\right]}{\min\left(V\left[A\right],V\left[B\right]\right)+V\left[C\right]},
\label{eq:overlap}
\end{equation}
where $V[\cdot]$ is the volume of regions $A,B,C$ defined as in Fig.~\ref{fig:overlap}. That is, with $I(x,y)=\mbox{Past}(x) \cap \mbox{Future}(y)$ denoting the Alexandrov interval between $x$ and $y$,
\begin{align}
  A &= I(a,c) \setminus I(b,c), \label{eq:A}\\
  B &= I(b,c) \setminus I(a,c), \label{eq:B}\\
  C &= I(a,c) \cap I(b,c).      \label{eq:C}
\end{align}
With this definition, the causal overlap is always in the range $\mathcal{O}(a,b) \in[0,1]$. If events $a$ and $b$ are time-like or null separated, then $\mathcal{O}(a,b) =1$, whereas for space-like separated events $\mathcal{O}(a,b) < 1$ unless both events are the same event.

The definition in Eq.~\eqref{eq:overlap} is valid for any Lorentzian manifold. Henceforth, we restrict our attentions only to the case when the spacetime is the Minkowski spacetime $\mathbb{M}^{d+1}$ of any dimension~$d$, and discuss other spacetimes in Section~\ref{sec:conclusions}. Given the freedom in the choice of event $c$, it is always possible to chose $c$ such that $V\left[A \cup C\right]=V\left[B \cup C\right]$. In $\mathbb{M}^{d+1}$, this is equivalent to saying that events $a$ and $b$ are at the same proper time $\tau_c$ from $c$ so that $V\left[A \cup C\right]=V\left[B \cup C\right] =v_d \tau_c^{d+1}$. If, without loss of generality, we set event $c$ and the origin of coordinates, then events $a$ and $b$ are located on the hyperboloid of constant proper time $\tau_c$ from event $c$.
Therefore, without loss of generality, we henceforth choose a reference frame in which events $a$ and $b$ are simultaneous. With these setting, events $a$ and $b$ are separated by the hyperbolic distance $d_{\mathbb{H}^{d}}(a,b)$, which is related to the distance in $\mathbb{M}^{d+1}$ as
\begin{equation}
d_{\mathbb{M}^{d+1}}(a,b)=2 \tau_c \sinh{\left( \frac{d_{\mathbb{H}^{d}}(a,b)}{2\tau_c}\right)}.
\end{equation}
The crucial point to notice here is that the causal overlap, as defined in Eq.~\eqref{eq:overlap}, is just a function of $\frac{d_{\mathbb{H}^{d}}(a,b)}{2\tau_c}$:
\begin{equation}
\mathcal{O}_{\mathbb{M}^{d+1}}(a,b)=f_d\left( \frac{d_{\mathbb{H}^{d}}(a,b)}{2\tau_c}\right),
\end{equation}
where $f_d(\cdot)$ is a function that depends on the spatial dimension of $\mathbb{M}^{d+1}$. The distance can thus be evaluated as
\begin{equation}
d_{\mathbb{M}^{d+1}}(a,b)=2 \tau_c \sinh{\left(f_d^{-1}(\mathcal{O}_{\mathbb{M}^{d+1}}(a,b))\right)}.
\label{eq:distance2}
\end{equation}
In dimension $d=1$, function $f_1(\cdot)$ takes a simple exponential form so that the causal overlap can be written as
\begin{equation}
\mathcal{O}_{\mathbb{M}^{2}}(a,b)=\exp{\left( -\frac{d_{\mathbb{H}^{1}}(a,b)}{\tau_c}\right)}
\end{equation}
so that the distance between $a$ and $b$ takes the form
\begin{equation}
d_{\mathbb{M}^{2}}(a,b)=\tau_c \frac{1-\mathcal{O}_{\mathbb{M}^{2}}(a,b)}{\sqrt{\mathcal{O}_{\mathbb{M}^{2}}(a,b)}}.
\end{equation}
In arbitrary dimensions, the causal overlap takes a cumbersome expression (see Appendix~\ref{app:overlap} for an integral representation). However, it is easy to see that in the limit $\tau_c \gg d_{\mathbb{H}^{d}}(a,b)$ it behaves as 
\begin{equation}
\mathcal{O}_{\mathbb{M}^{d+1}}(a,b) \approx 1-c_d \frac{d_{\mathbb{H}^{d}}(a,b)}{2\tau_c} 
\label{eq:overlapapprox}
\end{equation}
where
\begin{equation}
c_d=\frac{d+1}{\sqrt{\pi}} \frac{\Gamma \left( \frac{d}{2}\right)}{\Gamma \left( \frac{d+1}{2}\right)}.
\end{equation}
In principle, by knowing the exact expression of function $f_d(\cdot)$, we can use any arbitrary point $c$ (at proper time $\tau_c$ to $a$ and $b$) to evaluate the proper distance between $a$ and $b$ through Eq.~\eqref{eq:distance2}. However, precisely because point $c$ (and so $\tau_c$) is arbitrary, by choosing one for which $\tau_c \gg d_{\mathbb{H}^{d}}(a,b)$, we can use the asymptotic expression for the causal overlap Eq.~\eqref{eq:overlapapprox} and write that
\begin{equation}
d_{\mathbb{M}^{d+1}}(a,b)=\frac{2}{c_d}\lim_{\tau_c \rightarrow \infty} \tau_c (1-\mathcal{O}_{\mathbb{M}^{d+1}}(a,b)).
\label{eq:distance3}
\end{equation}

The estimation of distances based on causal overlaps in Eqs.~(\ref{eq:overlap},\ref{eq:distance2},\ref{eq:distance3}) has a number of desirable properties. The obvious one is that, thanks to the number-volume correspondence, the definition of causal overlaps in spacetimes translates straightforwardly to causal sets: the volumes of regions $A,B,C$ in Eq.~\eqref{eq:overlap} become the numbers of elements in the corresponding sets in a causal set.
This implies that distance estimations are intrinsic to the causal set graph, without any reference to an embedding continuous manifold. Another interesting property of Eq.~\eqref{eq:distance3} is that the dependence on spacetime dimension appears only as a multiplicative constant. Therefore, even without knowing the actual value of $d$, we can estimate distances up to a conformal factor. Finally, we note that there is an infinite number of possible events $c$ giving rise to the estimation of proper distances. Thus, the proper distance between two events can be understood as a measure of the entanglement of both events with their common past.

We now focus on causal sets arising from Poisson point processes on $\mathbb{M}^{d+1}$, in which case the effectiveness of Eqs.~\eqref{eq:distance2} and \eqref{eq:distance3} to recover the continuum in the limit $\rho \rightarrow \infty$ depends on the statistical properties of the causal overlap in Eq.~\eqref{eq:overlap}, which we can rewrite in this case as
\begin{equation}
\mathcal{O}_{\mathcal{C}}(a,b) \equiv \frac{N[C]}{N[A]+N[C]},
\label{eq:overlapinc}
\end{equation}
where $N[C]$ is the number of events in region~$C$, and $N[A]$ the number of events in region~$A$ or $B$ defined in Eqs.~(\ref{eq:A}-\ref{eq:C}).
We note that $N[A]$ and $N[C]$ are random variables defined in disjoint regions, and so they are statistically independent. Using this fact, it is easy to prove that the average value of $\mathcal{O}_{\mathcal{C}}(a,b)$ is
\begin{equation}
\langle \mathcal{O}_{\mathcal{C}}(a,b) \rangle =\mathcal{O}_{\mathbb{M}^{d+1}}(a,b).
\label{overlap:equivalence}
\end{equation}
Thus, the causal overlap as measured on $\mathcal{C}$ is, on average, the same as the causal overlap in $\mathbb{M}^{d+1}$.

Beyond the average, we can estimate the relative statistical error of the causal overlap as~\footnote{To derive this expression, we use error propagation from Eq.~\eqref{eq:overlapinc} taking into account that $N_c(a,b)  \mbox{ and } \Delta N_c(a,b)$ are independent Poisson random variables.}
\begin{equation}
\frac{\delta \mathcal{O}_{\mathcal{C}}(a,b)}{\langle \mathcal{O}_{\mathcal{C}}(a,b) \rangle}=\frac{1-\langle \mathcal{O}_{\mathcal{C}}(a,b) \rangle}{\sqrt{\rho V[C]}}  \sqrt{1+\frac{V[C]}{V[A]}}.
\label{eq:relativeerror}
\end{equation}
This expression approaches zero when $\rho \gg 1$, even at the smallest scales. To see this, suppose that $a$ and $b$ are separated by a proper length of the order of the Planck scale, that is $d_{\mathbb{M}^{d+1}}(a,b) \approx \rho^{-\frac{1}{d+1}} \sim t_P$. Using now Eq.~\eqref{eq:overlapapprox} and the definition of causal overlap, we see that $1-\langle \mathcal{O}_{\mathcal{C}}(a,b) \rangle\sim \frac{V[A]}{ V[C]} \sim \frac{t_P}{\tau_c}$. Combining these scaling results with Eq.~\eqref{eq:relativeerror} we conclude that the relative error of the causal overlap between two events space-like separated by a Planck length scales as
\begin{equation}
\frac{\delta \mathcal{O}_{\mathcal{C}}(a,b)}{\langle \mathcal{O}_{\mathcal{C}}(a,b) \rangle} \sim \left(\frac{t_P}{\tau_c}\right)^{\frac{d+2}{2}},
\end{equation} 
which goes to zero when $t_P\rightarrow 0$. These results show that the continuum can be recovered by measuring causal overlaps with number of events instead of actual volumes provided that $t_P\rightarrow 0$.

Finally, the error in the estimation of the distance by Eq.~\eqref{eq:distance3} will have a contribution from the error in the causal overlap computed above and the one from the estimation of $\tau_c$ which, according to Eq.~\eqref{error_tau}, is of the order $\sim t_P^{1-\beta_d}$. Combining both results, we conclude that the estimation of the distance is accurate whenever $\tau_c \gg t_P^{1-\beta_d}$.

\section{Numerical experiments}
\label{sec:simulations}

We run extensive numerical simulations to test the accuracy of Eq.~\eqref{eq:distance2} in measuring distances between space-like separated events, both at long and short scales.

To do so, we sprinkle uniformly at random a finite number of events $N$ in a box of side length $1$ in $\mathbb{M}^{d+1}$, $d=1,2,3$, so that the density of space-time events is set to $\rho=N$. The details of this sprinkling as well as other simulation details can be found in Appendix~\ref{app:sprinkling}. We set two events $a$ and $b$ separated by proper distance $l$ located at coordinates $ x^{\mu}_a=(1,-l/2,0,0)$ and $x^{\mu}_b=(1,l/2,0,0)$ for $d=3$ and similarly for $d=1,2$. To perform long-scale simulations we fix $l$ to a given value while increasing the density of events $\rho$. Instead, in short-scale simulations, we set the distance to the minimum distance allowed by the discretization of spacetime, that is $l=\rho^{1/(d+1)}$, while increasing the density of events.

\subsection{Determination of event $c$}

Given the volume-number correspondence, and so the equivalence between causal overlaps measured in $\mathbb{M}^{d+1}$ and $\mathcal{C}$ (as stated in Eq.~\eqref{overlap:equivalence}), we could use Eq.~\eqref{eq:distance2} to measure any distance using only information in $\mathcal{C}$. However, to use Eq.~\eqref{eq:distance2} we must first find an event $c$ that is simultaneously at (arbitrary) proper time $\tau_c$ from events $a$ and $b$ using only the structure of the causal set. In our simulation setup, such events have coordinate $x=0$ and are located at the intersection of the past light cones of $a$ and $b$. In principle, given the discreteness of the the causal set, it is not possible to find such events, although it is always possible to find events that are arbitrarily close to $x=0$. To find them, we use a double filter method. 

{\bf First filter.} Equation~\eqref{error_tau} poses a resolution limit in the estimation of proper times in the causal set. Thus, we first pre-select events $c$ such that
\begin{equation}
|n_{\cal C}(c,a)-n_{\cal C}(c,b)|<\rho^{\frac{\beta_d}{d+1}},
\label{event:c0}
\end{equation}
where $n_{\cal C}(c,a)$ is the longest path in $\mathbb{G}_{\mathcal C}$ connecting events $a$ and $c$. In this way, we can say that, up to our resolution limit, such $c$ events are at the same proper distance to $a$ and $b$. We note that this filter requires the knowledge of the density of the Poisson point process and the dimension of the embedding spacetime. This information is not contained in the causal set. We use it only to speed up the numerical simulations, but next we introduce a second filter that relies only on information contained in the causal set.

{\bf Second filter.} For each selected event $c$ in the previous step, we measure the number of events in the Alexandrov set between $a$ and $c$, $N[A \cup C]$, and between $b$ and $c$, $N[B \cup C]$. In $\mathbb{M}^{d+1}$, if $c$ is exactly at the same proper time from $a$ and $b$, the difference $Z_c(a,b) \equiv N[A \cup C]-N[B \cup C]$ is a random variable with the zero mean and variance 
\begin{equation}
\sigma_{Z_c}^2=\langle N[A \cup C] \rangle+\langle N[B \cup C] \rangle-2\langle N[C]\rangle.
\end{equation}
Therefore, out of all events $c$ selected in the previous step, we only keep those that satisfy
\begin{equation}
|Z_c(a,b)| < \frac{1}{2}\sqrt{N[A \cup C]+N[B \cup C]}\sqrt{1-\mathcal{O}_{\mathcal{C}}(a,b)}.
\label{event:c}
\end{equation}
The prefactor $1/2$ in this last inequality is arbitrary and can be selected to gauge the error in the estimation of event $c$. We emphasize an important point here that this inequality uses only information in the causal set, without any reference to the embedding Minkowski spacetime.

The question is whether Eq.~\eqref{event:c} selects events that are arbitrarily close to $x=0$ in the limit $\rho \rightarrow \infty$. Let us consider an event $c$ with an offset in the $x$ coordinate of $\delta x$, so that $x^{\mu}=(t,\delta x,y,z)$. The coordinates $t$, $y$, and $z$ are such that if the event had $\delta x=0$ the proper time from $a$ and $b$ to $c$ would be $\tau_c$. Then, the expected value of $Z_c(a,b)$ when $\delta x \ne 0$ is
\begin{align}
\langle Z_c(a,b) \rangle=v_d \rho \tau_c^{d+1}&\left[\left(1-\frac{\delta x^2+l \delta x}{\tau_c^2}\right)^{\frac{d+1}{2}}\right.\\
&-\left.\left(1-\frac{\delta x^2-l \delta x}{\tau_c^2} \right)^{\frac{d+1}{2}}\right].\nonumber
\end{align}
By comparing this expression with Eq.~\eqref{event:c}, it is possible to determine the value of $\delta x$ above which event $c$ is rejected as a suitable event. Assuming that $\delta x/\tau_c \ll 1$, the event $c$ is rejected when
\begin{equation}
\frac{l \delta x}{\tau_c^2} > \sqrt{\frac{1-\mathcal{O}_{\mathcal{C}}(a,b)}{2 (d+1)^2 v_d \rho \tau_c^{d+1}}},
\label{reject:c}
\end{equation}
where we have used that $N[A \cup C]+N[B \cup C] \approx 2 v_d \rho \tau_c^{d+1}$. In the case of long-scale distances with $l$ fixed, the causal overlap is constant and the right hand side of inequality Eq.~\eqref{reject:c} scales as $t_P^{(d+1)/2}$. This implies that in the continuum limit $t_P \rightarrow 0$, the selection criteria of events $c$ is more and more stringent, with selected events approaching arbitrarily close to $x=0$. In the case of short-scale distances $l \propto t_P$, the causal overlap scales as $1-\mathcal{O}_{\mathcal{C}}(a,b) \sim t_P/\tau_c$, so that the inequality Eq.~\eqref{reject:c} becomes
\begin{equation}
\frac{\delta x}{\tau_c} > \left( \frac{t_P}{\tau_c} \right)^{\frac{d}{2}}.
\end{equation}
Again, we see that the right hand side of the inequality goes to zero in the continuum limit, so that even at the smallest length scales the selection of events $c$ becomes asymptotically exact.

Notice that, in fact, we could use only Eq.~\eqref{event:c} to select events $c$, which relies only on information in the causal set, whereas Eq.~\eqref{event:c0} uses information about the dimension of the embedding space. However, using the first step is more computationally efficient because, in this case, we only have to measure $N[A\cup C], N[B \cup C]$, and $N[C]$ for a subset of events $c$. 

\begin{figure}[t]
\centerline{\includegraphics[width=\columnwidth]{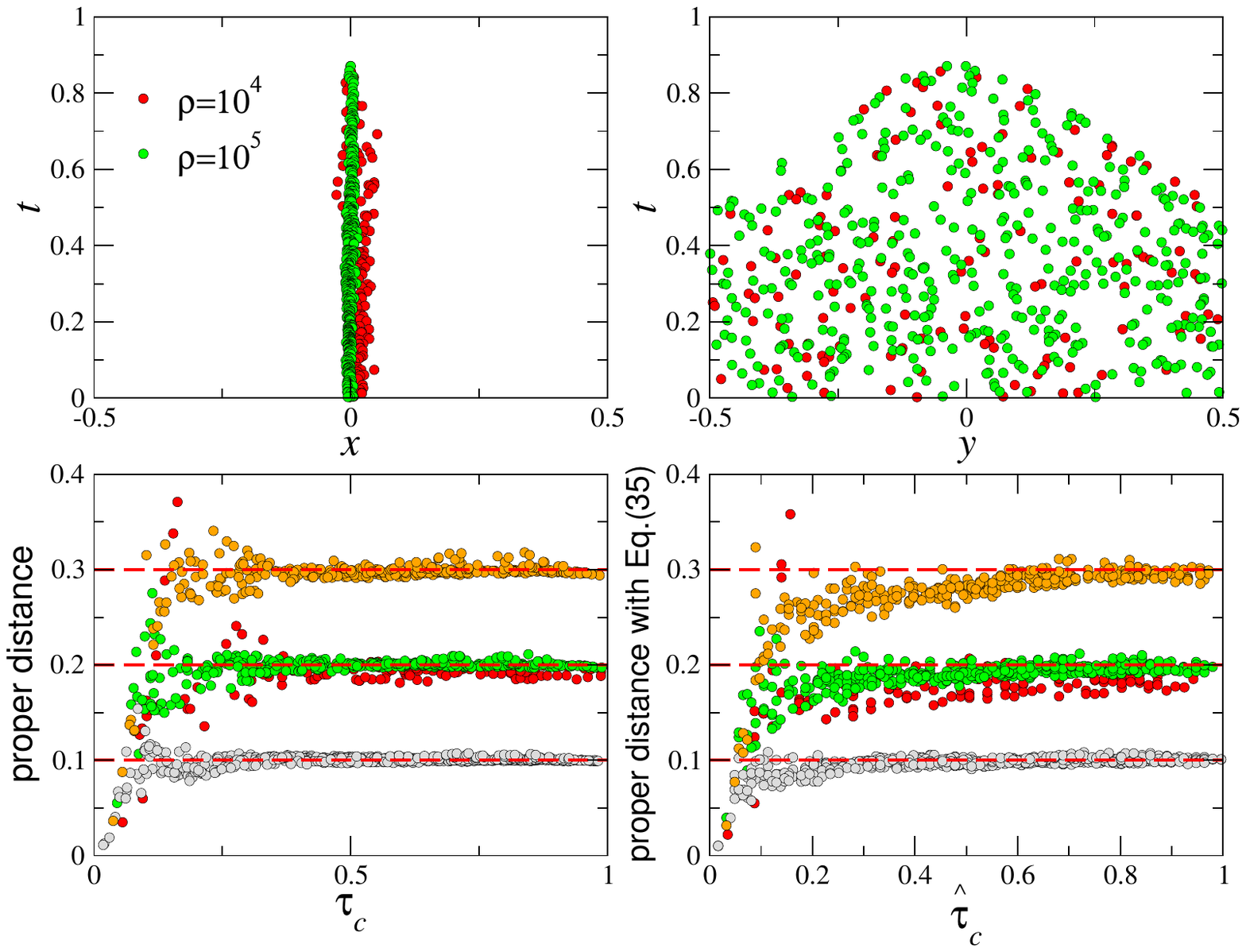}}
\caption{{\bf Long-scale simulations in $\mathbb{M}^3$.} The top row shows the $x$ (left) and $y$ (right) coordinates of selected events $c$ used to estimate the distance between events $a$ and $b$ for two different densities, $\rho=10^4$ (red circles) and $\rho=10^5$ (green circles). The proper distance between events $a$ and $b$ is $0.2$. The bottom left plot shows the distance estimations for all selected events $c$ shown in the top plots as a function of the actual proper time $\tau_c$ of event $c$. The orange and gray circles correspond to distance estimations of events separated by proper distances $0.1$ (gray) and $0.3$ (orange) at density $\rho=10^5$. The dashed lines indicate the actual distances that are inferred. The bottom right plot shows the same as the left plot but using the estimation $\hat{\tau}_c$ of proper times of events $c$ in Eq.~\eqref{eq:tauestimate}.
}
\label{fig:longM3}
\end{figure}

\begin{figure}[th]
\centerline{\includegraphics[width=\columnwidth]{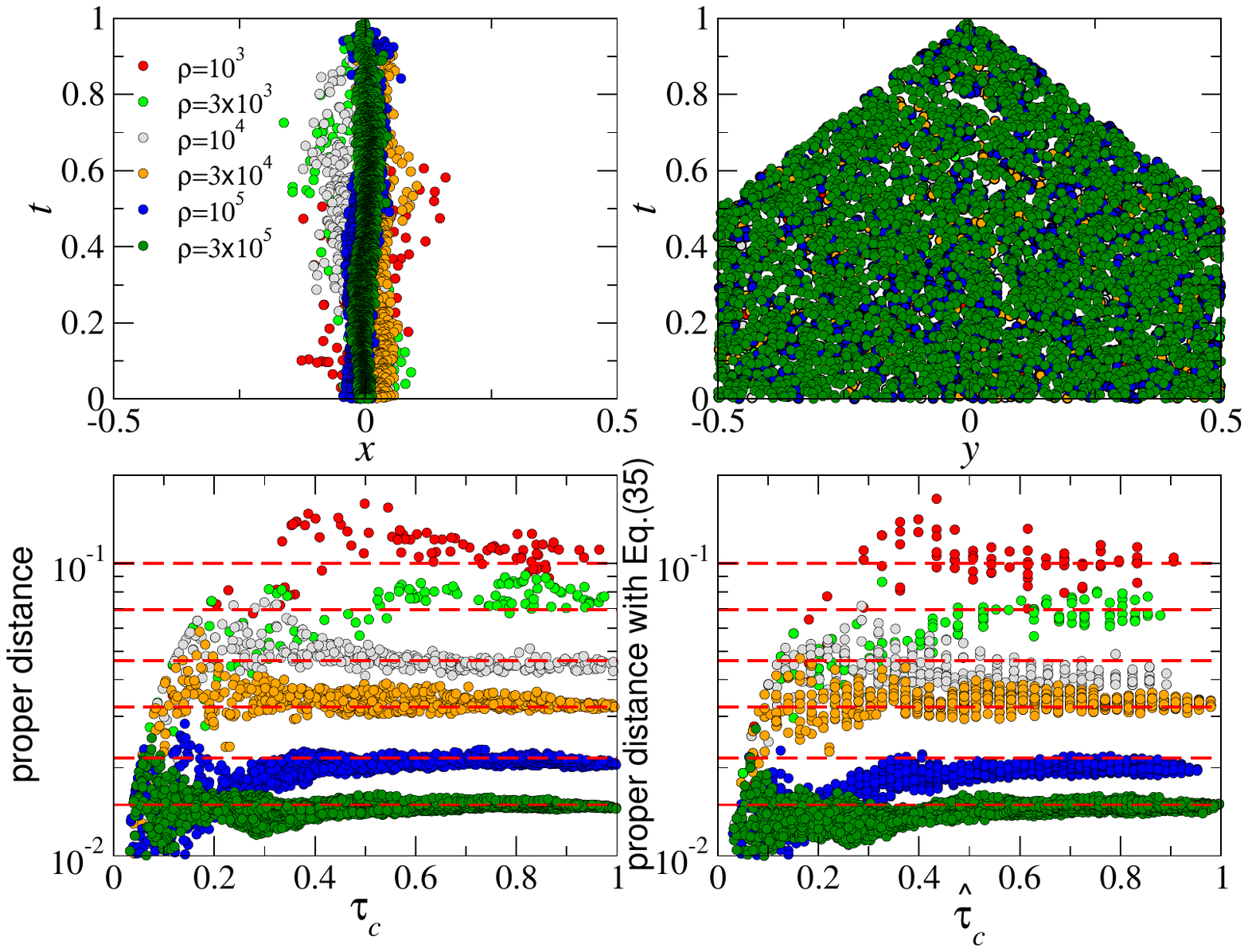}}
\caption{{\bf Short-scale simulations in $\mathbb{M}^3$.} The top row shows the $x$ (left) and $y$ (right) coordinates of selected events $c$ used to estimate the distance between events $a$ and $b$, which is set to $l=\rho^{-1/(d+1)}$ with $\rho=10^3,\cdots,3\times10^5$. The bottom left plot shows the distance estimations for all events shown in the top plots as a function of the proper times $\tau_c$ of events~$c$. The dashed lines indicate the actual distances that are inferred. The bottom right plot shows the same as the left plot but using the estimation $\hat{\tau}_c$ of proper times of events $c$ in Eq.~\eqref{eq:tauestimate}.}
\label{fig:shortM3}
\end{figure}

\subsection{Numerical results}

Figure~\ref{fig:longM3} shows simulation results for long distance estimations in $\mathbb{M}^3$. The top row in Fig.~\ref{fig:longM3} shows the $x$ (left) and $y$ (right) coordinates of selected events $c$ at two different densities, $\rho=10^4$ and $\rho=10^5$. As it can be seen, selected events are more concentrated near the plane $x=0$ when the density is increased, as predicted in the previous section. For each selected event $c$, we evaluate the proper distance between events $a$ and $b$ using the numerical solution of Eq.~\eqref{eq:distance2}, where $\tau_c$ is computed as $\tau_c=(\tau_{{\mathbb M}^{d+1}}(c,a)+\tau_{{\mathbb M}^{d+1}}(c,b))/2$ using the actual coordinates in $\mathbb{M}^{d+1}$ of events $a$, $b$, and $c$. In this way, the randomness in the estimation of distances comes from the fluctuations associated to the causal overlap alone.

Concerning event $c$, a priori, any such event can be used in Eq.~\eqref{eq:distance2}. However, those with low proper time $\tau_c \lesssim l/2$ have a higher statistical error due to the spacetime discretization. Besides, due to the simulation setup, events with values of the $y$ coordinate far from $y=0$ have part of their past light cones outside the simulated box, inducing an extra error term. This problem can be, however, minimized by choosing events with $\tau_c \approx 1$ as such events have necessarily $y \approx 0$.
The bottom left plot in Fig.~\ref{fig:longM3} shows the inferred proper distances when $l=0.1,0.2,0.3$ as a function of the actual proper time $\tau_c$ of all selected events $c$ compared against the actual values indicated by the dashed red lines. We can see that the error in the estimation of distances due to the choice of events $c$ is small and becomes even smaller when the density increases. However, in this result we still use a bit of information not contained in the causal set because we plug the actual proper times $\tau_c$ of events $c$ to check our predictions. Instead, the bottom right plot in Fig.~\ref{fig:longM3} shows the same inferred proper distances but estimating the value of $\tau_c$ using only the causal set structure:
\begin{equation}
\hat{\tau}_c=\frac{1}{2}\alpha_d \rho^{-1/(d+1)}(n_{\cal C}(c,a)+n_{\cal C}(c,b)),
\label{eq:tauestimate}
\end{equation}
with $\alpha_d$ measured numerically in the simulations. In this case, the estimation of proper distances contains two sources of stochasticity, the one associated with causal overlaps and the one associated to the estimation of $\tau_c$. However, given that estimations of proper times in causal sets can be done with a very small error when $\tau_c \gg t_P$, as in the present case, we observe very similar results as on the plot on the left for $\tau_c >0.5$. 

\begin{figure}[th]
\centerline{\includegraphics[width=\columnwidth]{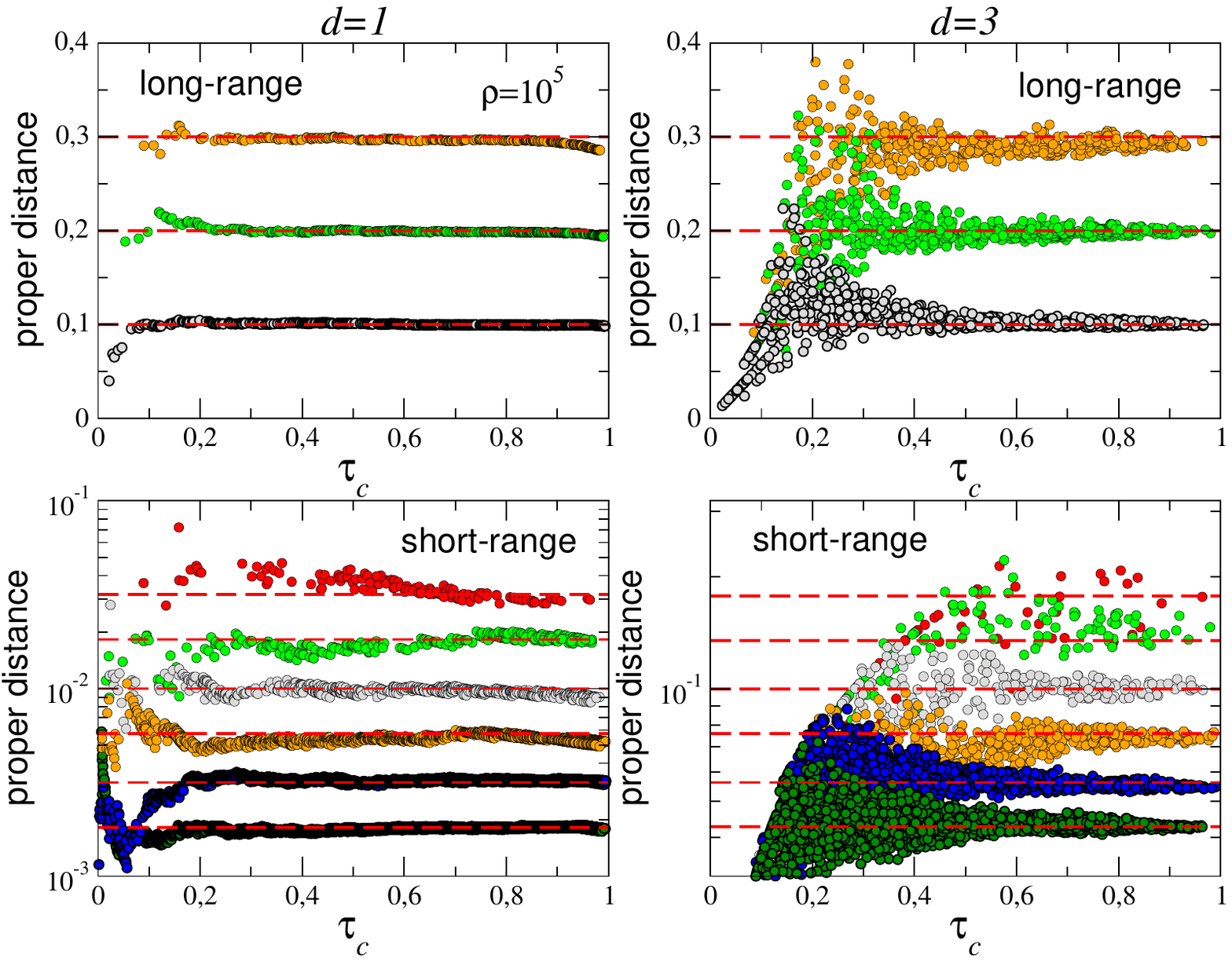}}
\caption{{\bf Long and Short-scale simulations in $\mathbb{M}^2$ and $\mathbb{M}^4$ as a function of the proper time of event $c$.} The left column shows results for long-scale (top) and short-scale (bottom) simulations in $\mathbb{M}^2$ and the right column for $\mathbb{M}^4$. For the long-scale simulations, we fix the density to $\rho=10^5$. In the case of short-scale simulations, we set the density to $\rho=10^3,3\times 10^3,10^4,3\times  10^4,10^5,3\times  10^5$ (from top to bottom) and measure the corresponding distances, which are highlighted by the red dashed lines.}
\label{fig:M2M4}
\end{figure}

Figure~\ref{fig:shortM3} shows the same analysis but for the estimation of short distances. In particular, we set $l=\rho^{-1/(d+1)}$ and increase $\rho$ from $10^3$ to $3 \times10^5$. As in the case of long-scale distances, selected events $c$ are more aligned along the plane $x=0$ as the density increases. The plots at the bottom of Fig.~\ref{fig:shortM3} show the perfect estimation of all distances at any value of the density, showing that, indeed, distances can be measured in causal sets at all scales. Finally, Fig.~\ref{fig:M2M4} shows results for $\mathbb{M}^2$ and $\mathbb{M}^4$ Minkowski spacetimes, showing that proper distances can be measured with the same method in any dimension. Notice that the differences in magnitude of the fluctuations in different dimensions are due to the fact that we use the same density $\rho$ for $d=1,2,$ and $d=3$, which result in different values of $t_P$ in different dimensions. 

\section{Kinematics of Minkowski causal sets}
\label{sec:kinematics}

The continuum limit of causal set theory suggests that at scales exceeding the Planck scale, it becomes feasible to reconstruct the spacetime manifold structure. This reconstruction includes the capability to define inertial frames of reference directly on the causal set, using only the information contained in it. Consequently, events within a causal set can be described using spacetime coordinates. Such inertial frames play a pivotal role: in addition to the spatial-temporal characterization of events, they also enable the measurement of velocities of time-like trajectories within a causal set, thereby establishing the kinematics intrinsic to causal sets.

Our methodology facilitates the accurate measurement of proper times and distances between events that are either time-like or space-like separated using information in the causal set alone. This accuracy allows for a reliable definition of inertial frames of reference, leading to a clearer understanding of instantaneous velocities along time-like curves. This approach enhances our ability to interpret and analyze the dynamics within causal sets, shedding light on the complex interplay between discrete and continuous visions of spacetime.

In this section we discuss two aspects of this program: how such references frames can be set up, and how some kinematic aspects enabled by them (Lorentz factor) can be measured.

\subsection{Reference frames}
In general, an inertial frame of reference can be defined by a geodesic time-like curve, with one of the events in the geodesic chosen as its origin of coordinates. In the causal set, a geodesic time-like curve between two events is defined as the longest chain of links connecting both events.

Consider the geodesic made of the sequence of ordered events 
\begin{equation}
\mathbb{A}=\{a_i\in \mathcal{C} | \cdots \prec a_{i-1}\prec a_i\prec a_{i+1}\prec \cdots \}. 
\end{equation}
Given an event $b$ not contained in $\mathbb{A}$, there is a finite number of events in $\mathbb{A}$ that are space-like separated from $b$, that we denote by $S(b,\mathbb{A})$. We then define the distance between event $b$ and geodesic $\mathbb{A}$ as
\begin{equation}
d(\mathbb{A},b)=\sup_{a_i \in S(b,\mathbb{A})} d_{\mathbb{M}^{d+1}}(a_i,b)=d_{\mathbb{M}^{d+1}}(a_m,b),
\end{equation}
with $a_m$ as the event in $S(b,\mathbb{A})$ maximizing the proper distance to $b$. If $a_0 \in \mathbb{A}$~\footnote{Without loss of generality, we assume $a_0 \prec a_m$} is chosen as the origin of coordinates, we can define the space-time coordinates of event $b$, $x_b^\mu =(x_b^0,\mathbf{x}_b)$,
in the reference frame defined by $\mathbb{A}$ and $a_0$ as
\begin{equation}
x_b^0=\tau_{\mathcal{C}}(a_0,a_m) \; , \; |\mathbf{x}_b|=d_{\mathbb{M}^{d+1}}(a_m,b),
\end{equation}
where $\tau_{\mathcal{C}}(a_0,a_m)$ is the proper time from $a_0$ to $a_m$ measured in $\mathcal{C}$, as given by Eq.~\eqref{eq:1}. Again, notice that this definition is intrinsic to the causal set graph because $d_{\mathbb{M}^{d+1}}(a_m,b)$ is measured in terms of causal overlaps in $\mathcal{C}$ and $\tau_{\mathcal{C}}(a_0,a_m)$ is proportional to the number of steps in $\mathbb{A}$ from $a_0$ and $a_m$. 

The determination of the individual components of the spatial part of $b$'s coordinates can be performed by defining the subspace of simultaneous events to $a_m$ in $\mathbb{A}$, $\mathbb{B}_{\bot}(\mathbb{A})$, defined as the set of space-like separated events $\{b_{\bot}\in \mathcal{C}\}$ that have $a_m$ as the event in $\mathbb{A}$ maximizing the distance $d_{\mathbb{M}^{d+1}}(a_m,b_{\bot})$. The set $\mathbb{B}_{\bot}(\mathbb{A})$ is a numerable set of space-like separated events. Thus, using Eq.~\eqref{eq:distance3}, we can compute the matrix of proper lengths among them. Using this matrix, the dimension of the subspace can then be easily estimated by measuring the volume of balls as a function of their radius, similar to the Hausdorff and spectral dimension estimations proposed in~\cite{Eichhorn_2019,Eichhorn_2019b}. Our approach represents an alternative to the Myrheim-Meyer~\cite{Meyer:1988} and midpoint-scaling~\cite{Mid:point} methods to estimate the spacetime dimension~\cite{Reid:2003kc}. Finally, we can use any embedding method from the computer science literature (like Laplacian Eigenmaps~\cite{Belkin:2003aa}) that, by using the matrix of distances, finds a mapping between events from $\mathbb{B}_{\bot}(\mathbb{A})$ and points in $\mathbb{R}^d$. This program will be developed in a forthcoming publication.

\subsection{Lorentz factor}

Beyond the spatial-temporal characterization of events in the reference frame defined above, we can also characterize how objects move relative to this frame. Using the results in the previous section, we can make a step forward and measure the instantaneous velocity of a time-like curve and the corresponding Lorentz factor.

Suppose that a given observer is at event $b$ and travels to its future event $b'$ using a geodesic path. This new event $b'$ is at distance $d(\mathbb{A},b')$ to $\mathbb{A}$ with a corresponding event $a_m'$ in $\mathbb{A}$. Suppose that, during this transition, the observer's proper time increases by $n_{\cal C}(b,b')$ steps in the causal set. Therefore, the Lorentz factor --defined as the ratio between the variation of the coordinate time and proper time-- can be defined in the causal set as
\begin{equation}
\gamma_{\mathbb{A}}=\frac{n_{\cal C}(a_m,a_m')}{n_{\cal C}(b,b')}.
\label{lorentz:gamma}
\end{equation}
Using this equation, we can derive an expression for the speed of a time-like curve from $b$ to $b'$ in the reference frame defined by $\mathbb{A}$, $v_\mathbb{A}^2=1-\gamma_{\mathbb{A}}^{-2}$. Similarly, the radial velocity $\dot{r}_{\mathbb{A}}$ can be computed as the variation in the distance to the geodesic $\mathbb{A}$, that is,
\begin{equation}
\dot{r}_{\mathbb{A}}=\frac{d(\mathbb{A},b')-d(\mathbb{A},b)}{\tau_{\mathcal{C}}(a_m,a_m')}.
\end{equation} 
And using this expression along with Eq.~\eqref{lorentz:gamma}, we can evaluate the modulus of the angular component of the velocity.

In the limit $t_P \rightarrow 0$ the expression for the Lorentz factor in Eq.~\eqref{lorentz:gamma} converges to the actual value of $\gamma$ in $\mathbb{M}^{d+1}$ while still describing an infinitesimal variation of the time-like curve. This can be achieved when the total proper time between $b$ and $b'$ is very small, that is when $n_{\cal C}(b,b')t_P \ll 1$ and, simultaneously, the relative error in the estimation of proper times is also very small. According to Eq.~\eqref{error_tau}, this condition is fulfilled as long as $n_{\cal C}(b,b')t_P^{\beta_d} \gg 1$. This defines a range in the number of steps in $\mathcal{C}$ 
\begin{equation}
t_P^{-\beta_d} \ll n_{\cal C}(b,b') \ll t_P^{-1}
\label{eq:range}
\end{equation}
within which the accuracy in the evaluation of $\gamma$ using Eq.~\eqref{lorentz:gamma} is high while the proper time between events $b$ and $b'$ is small. Since $\beta_d<1$, in the limit $t_P \rightarrow 0$ the upper bound in Eq.~\eqref{eq:range} grows faster than the lower limit so that it is always possible to measure instantaneous velocities of time-like curves (not necessarily geodesic) in any reference frame with arbitrary precision.

\begin{figure}[t]
\centerline{\includegraphics[width=\columnwidth]{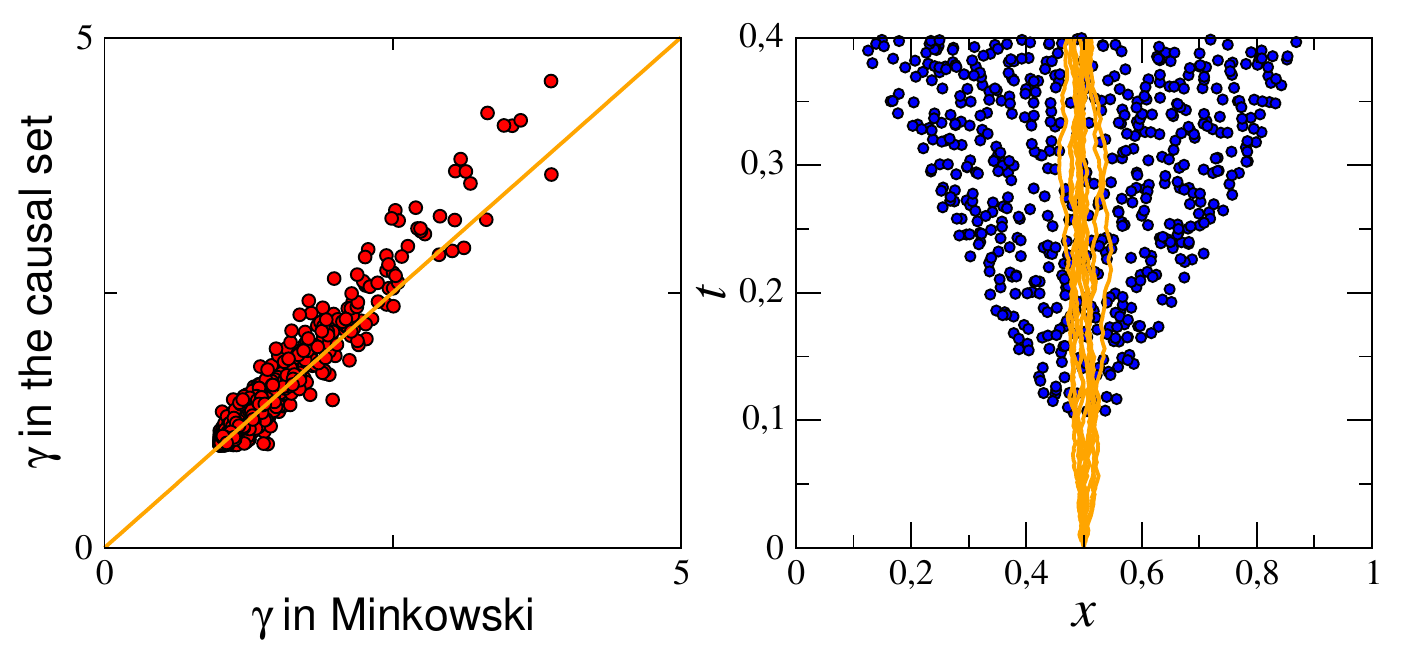}}
\caption{{\bf Measuring the Lorentz factor in the causal set.} The plot on the right shows the geodesics $\mathbb{A}$ used as a reference frame (orange) and the set of events $b'$ used to evaluate $\gamma_{\mathbb{A}}$. The plot on the left shows $\gamma_{\mathbb{A}}$ from Eq.~\eqref{lorentz:gamma} versus the actual value in $\mathbb{M}^2$. The density is set to $\rho=10^5$.}
\label{fig:lorentz}
\end{figure}

We perform numerical simulations to test Eq.~\eqref{lorentz:gamma} in~$\mathbb{M}^2$. We generate ten different realizations of the Poisson sprinkling in $\mathbb{M}^2$ at density $\rho=10^5$. For each realization, we set the event $a_0$ to be at the origin, $a_0=(0,0)$, and find the geodesic $\mathbb{A}$ connecting $a_0$ to the event closest to $(0.5,0)$. These geodesics are depicted by the orange lines in Fig.~\ref{fig:lorentz}. We then set events $b$ to be also at the origin, $b=a_0$, and sample uniformly at random $100$ events $b'$ per sprinkling in the domain $t(b') \in (0,0.5)$, $x(b') \in (0.1,0.9)$, and $n_{\cal C}(b,b')>t_P^{-\beta_d}$, which are also shown in Fig.~\ref{fig:lorentz}. These settings limit the maximum sampled value of the Lorentz factor to $\gamma <5$ (close to $99\%$ of the speed of light). For each sampled event~$b'$, we compute $\gamma_{\mathbb{A}}$ using Eq.\eqref{lorentz:gamma}. In Fig.~\ref{fig:lorentz}, we compare this estimation to the actual value of $\gamma$ measured with the actual coordinates in $\mathbb{M}^2$. Despite the noise in the estimation due to finite size effects, the agreement is very good. We notice, however, a small systematic bias in $\gamma_{\mathbb{A}}$ which has two different sources. The first one is due to Jensen's inequality, which implies that $\langle n_{\cal C}(b,b')^{-1} \rangle > \langle n_{\cal C}(b,b') \rangle^{-1}$. The second source of the bias are the intervals, used in the computation of causal overlaps, that partially lie outside of the simulation box, thereby affecting the estimation of proper distances. This effect is more prominent for events with high values of $t$ and $x$, and so with high values of the Lorentz factor.

\section{Conclusion}
\label{sec:conclusions}

The estimation of distances in causal sets is a fundamental roadblock on the route towards understanding their continuum limits. Here we introduced a methodology to measure spatial distances in causal sets. This methodology works well all the way down to the Planck scale, the ultimate granularity of spacetime structure. This breakthrough advances our understanding of the continuum limit of discrete spacetimes, and opens avenues for defining local reference frames~\cite{Glaser:2013hb} and studying kinematics in them.

While our findings are anchored in Minkowski spacetimes, it is possible to extend them to other spacetimes. Indeed, the metric tensor of any Lorentzian manifold can be approximated by the Minkowski metric to the first order in a small neighborhood around any point, making local physics indistinguishable from that in the flat Minkowski spacetime~\cite{wald2010general}. In addition to showing that proper distances among space-like separated events can be measured all the way down to the Planck scale, we have also showed that to get accurate estimates for such distances, the proper time $\tau_c$ to events $c$ must be greater than $t_P^{1-\beta_d}$, where $\beta_d$ is from Eq.~\eqref{error_tau}. This limit sets the minimum scale above which a local neighborhood can be defined around a given event. Therefore, if the characteristic scale of the curvature of a spacetime is larger than $t_P^{1-\beta_d}$, our approach can be used to define infinitesimal distances around any event in such a spacetime. In addition to that, the number of events within such neighborhoods can be determined, thus effectively defining the metric tensor.

Central to our approach is the definition of causal overlaps between events. These overlaps are a form of entanglement within the shared past of the events, a picture applicable to any spacetime. We believe that this form of entanglement must play a pivotal role in the ambitious goal of constructing models of evolving causal sets in which spatial geometry emerges from their dynamics~\cite{rideout1999,Varadarajan:2006lq,Ahmed:2010vn,Surya_2020,Bento_2022,Zalel2023}. Such geometrogenesis models may not use anything related to continuous spacetimes, since there are no continuous spacetimes beyond the Planck scale to begin with. The causal set evolution rules in such models are allowed to use only the structure of the causal sets that they grow. Yet when grown to large continuum-limit sizes, these causal sets must be indistinguishable at a large scale from causal sets obtained by sprinkling onto continuous spacetimes with spatial geometry of our physical spacetime. We believe that it is very difficult, nearly impossible really, to build such models without good understanding of how spatial distances can be reliably estimated in causal sets using only their structure.

\section{Acknowledgments}
M.B.\ acknowledges support from grant TED2021-129791B-I00 funded by MCIN/AEI/10.13039/501100011033 and the ``European Union NextGenerationEU/PRTR'', Grants PID2019-106290GB-C22 and PID2022-137505NB-C22 funded by MCIN/AEI/10.13039/501100011033; Generalitat de Catalunya grant number 2021SGR00856. M.B.\ acknowledges the ICREA Academia award, funded by the Generalitat de Catalunya. D.K.\ acknowledges NSF Grant Nos.\ IIS-1741355 and CCF-2311160.\\

\appendix

\section{Causal overlap in $\mathbb{M}^{d+1}$}
\label{app:overlap}

Without loss of generality, we place event $c$ at the origin of coordinates and events $a$ and $b$ at $(t_a, x_{1,a},\vec{0})$ and $(t_b, x_{1,b},\vec{0})$, respectively, with $t_a=t_b$ and $x_{1,b}=-x_{1,a}$. Events $a$ and $b$ are at proper time $\tau_c$ from $c$ and so  $t_a^2=\tau_c^2+x_{1,a}^2$. To compute the causal overlap $\mathcal{O}_{\mathbb{M}^{d+1}}(a,b)$, we first must compute the volume $V_c(a,b)$ at the intersection of the future light cone of $c$, given by the equation $t=r$, and the past light cones of $a$ and $b$, which in spherical coordinates are given by 
\begin{equation}
t=t_a-\sqrt{r^2+x_{1,a}^2-2x_{1,a}r \cos{\theta}},
\end{equation}

\begin{equation}
t=t_b-\sqrt{r^2+x_{1,b}^2+2 x_{1,b}r \cos{\theta}},
\end{equation}
and where we have chosen the coordinate $x_1=r \cos{\theta}$, with $\theta \in[0,\pi]$. After some algebra, we obtain
\begin{widetext}
\begin{equation}
V_c(a,b)=\frac{4 \pi^{(d-1)/2}}{\Gamma\left( \frac{d-1}{2}\right)} \int_0^{\pi/2} \sin^{d-2}\theta d\theta \int_0^{r^*(\theta)} r^{d-1} \left(t_a-r-\sqrt{r^2+x_{1,a}^2+2x_{1,a}r \cos{\theta}} \right)dr
\end{equation}
where
\begin{equation}
r^*(\theta)=\frac{\tau_c^2}{2(t_a+x_{1,a} \cos{\theta})}
\end{equation}
Notice that by setting $x_{1,a}=0$, we recover the volume of the Alexandrov set in Eq.~\eqref{alexandrov}. Thus, the causal overlap can be written as
\begin{equation}
\mathcal{O}_{\mathbb{M}^{d+1}}(a,b)=\frac{2^{d+2}(d+1)\Gamma\left( 1+\frac{d}{2} \right)}{\sqrt{\pi}\Gamma\left( \frac{d-1}{2}\right)} \int_0^{\pi/2} \sin^{d-2}\theta d\theta \int_0^{\hat{r}^*(\theta)} r^{d-1} \left(\sqrt{1+\hat{x}_{1,a}^2}-r-\sqrt{r^2+\hat{x}_{1,a}^2+2\hat{x}_{1,a}r \cos{\theta}} \right)dr
\label{eq:overlap2}
\end{equation}
\end{widetext}
where
\begin{equation}
\hat{r}^*(\theta)=\frac{1}{2(\sqrt{1+\hat{x}_{1,a}^2}+\hat{x}_{1,a} \cos{\theta})} \; \; \mbox{and}  \; \; \hat{x}_{1,a}=\frac{x_{1,a}}{\tau_c}
\end{equation}
Taking the limit $\hat{x}_{1,a}\rightarrow 0$ in Eq.~\eqref{eq:overlap2} leads to the asymptotic result Eq.~\eqref{eq:overlapapprox}.
 
\section{Poisson point processes and other simulation details}
\label{app:sprinkling}

A Poisson point process~\cite{last2017lectures} of rate~$\delta$ in an arbitrary manifold is defined as a random point process in which the random number of points~$n$ in any chunk of space of volume $V$ is given by the Poisson distribution with mean~$\delta V$:
\begin{equation}
\text{Prob}\{ n  | \delta, V \}=\frac{1}{n!} (\delta V)^n e^{-\delta V},
\end{equation}
and the numbers of points in any two nonintersecting volumes are independent random variables. If the manifold is Minkowski $\mathbb{M}^{d+1}$, then
the Lorentz invariance of volumes guarantees that Poisson sprinklings are Lorentz invariant as well.

To generate Poisson sprinklings of rate~$N$ in $\mathbb{M}^{d+1}$, we place $N$ events at random spacetime positions $x^\mu=(t,\mathbf{x})$ as follows.
\begin{enumerate}
\item
Each of the $d$ spatial coordinates $\mathbf{x}_i$ of each event $i=1,\ldots,N$ is chosen uniformly at random within the interval $(-\frac{1}{2},\frac{1}{2})$.
\item
The temporal coordinate of event~$i$ is set to
\begin{equation}
t_i=t_{i-1}+\xi_i,
\end{equation}
where $t_0=0$ and $\xi_i>0$ are random numbers drawn independently from the exponential distribution of rate~$N$ whose PDF is $p(\xi)=Ne^{-N\xi}$.
\end{enumerate}
In assigning the temporal coordinates in the last equation, we rely on the fact that time differences between any two consecutive events in a Poisson point process are independent exponentially distributed random variables~\cite{last2017lectures}. This time-ordered coordinate assignment is also convenient for simulation purposes because to check whether an event~$j$ is to the future or past of an event~$i$, we only need to check whether $j>i$ or $j<i$, respectively. An alternative to that, with random time coordinate assignments, is to keep link directions below, which would be more computationally intensive. We note that the time coordinate of the $N$'th point (the height of our hypercube) is the sum of $N$ exponentially distributed random numbers $\xi_i$ of rate~$N$, so it is also a random number whose mean is $1$ and whose distribution is a Gamma distribution.

Once every event within our simulation hypercube has its random coordinates in $\mathbb{M}^{d+1}$ assigned as prescribed by the sprinkling procedure above, a causal relation between all pairs of events $i$ and $j$ is defined whenever $|x_i^\mu-x_j^\mu|<0$, with $j$ in the future of $i$ if $j>i$ and in its past otherwise. The set of events combined with their causal relations defines the causal set $\mathcal{G}$. We then apply a transitive reduction to the causal set $\mathcal{G}$ so that only irreducible links remain in its graph representation~$\mathbb{G}_{\mathcal{G}}$.

The actual spacetime coordinates of the events are then \emph{not} used in the distance estimation, but they \emph{are} used to compare the results of this estimation against the true distances. The dimension of spacetime is used in the distance estimation Eq.~\eqref{eq:distance2}.


\begin{thebibliography}{48}%
\makeatletter
\providecommand \@ifxundefined [1]{%
 \@ifx{#1\undefined}
}%
\providecommand \@ifnum [1]{%
 \ifnum #1\expandafter \@firstoftwo
 \else \expandafter \@secondoftwo
 \fi
}%
\providecommand \@ifx [1]{%
 \ifx #1\expandafter \@firstoftwo
 \else \expandafter \@secondoftwo
 \fi
}%
\providecommand \natexlab [1]{#1}%
\providecommand \enquote  [1]{``#1''}%
\providecommand \bibnamefont  [1]{#1}%
\providecommand \bibfnamefont [1]{#1}%
\providecommand \citenamefont [1]{#1}%
\providecommand \href@noop [0]{\@secondoftwo}%
\providecommand \href [0]{\begingroup \@sanitize@url \@href}%
\providecommand \@href[1]{\@@startlink{#1}\@@href}%
\providecommand \@@href[1]{\endgroup#1\@@endlink}%
\providecommand \@sanitize@url [0]{\catcode `\\12\catcode `\$12\catcode
  `\&12\catcode `\#12\catcode `\^12\catcode `\_12\catcode `\%12\relax}%
\providecommand \@@startlink[1]{}%
\providecommand \@@endlink[0]{}%
\providecommand \url  [0]{\begingroup\@sanitize@url \@url }%
\providecommand \@url [1]{\endgroup\@href {#1}{\urlprefix }}%
\providecommand \urlprefix  [0]{URL }%
\providecommand \Eprint [0]{\href }%
\providecommand \doibase [0]{http://dx.doi.org/}%
\providecommand \selectlanguage [0]{\@gobble}%
\providecommand \bibinfo  [0]{\@secondoftwo}%
\providecommand \bibfield  [0]{\@secondoftwo}%
\providecommand \translation [1]{[#1]}%
\providecommand \BibitemOpen [0]{}%
\providecommand \bibitemStop [0]{}%
\providecommand \bibitemNoStop [0]{.\EOS\space}%
\providecommand \EOS [0]{\spacefactor3000\relax}%
\providecommand \BibitemShut  [1]{\csname bibitem#1\endcsname}%
\let\auto@bib@innerbib\@empty
\bibitem [{\citenamefont {Planck}(1899)}]{planck1899}%
  \BibitemOpen
  \bibfield  {author} {\bibinfo {author} {\bibfnamefont {M.}~\bibnamefont
  {Planck}},\ }\href@noop {} {\bibfield  {journal} {\bibinfo  {journal}
  {Proceedings of the Royal Prussian Academy of Sciences}\ }\textbf {\bibinfo
  {volume} {5}},\ \bibinfo {pages} {479} (\bibinfo {year} {1899})}\BibitemShut
  {NoStop}%
\bibitem [{\citenamefont {Heisenberg}(1925)}]{heisenberg1925}%
  \BibitemOpen
  \bibfield  {author} {\bibinfo {author} {\bibfnamefont {W.}~\bibnamefont
  {Heisenberg}},\ }\href@noop {} {\bibfield  {journal} {\bibinfo  {journal}
  {Zeitschrift f{\"u}r Physik}\ }\textbf {\bibinfo {volume} {33}},\ \bibinfo
  {pages} {879} (\bibinfo {year} {1925})}\BibitemShut {NoStop}%
\bibitem [{\citenamefont {Schwarzschild}(1916)}]{schwarzschild1916}%
  \BibitemOpen
  \bibfield  {author} {\bibinfo {author} {\bibfnamefont {K.}~\bibnamefont
  {Schwarzschild}},\ }\href@noop {} {\bibfield  {journal} {\bibinfo  {journal}
  {Proceedings of the Royal Prussian Academy of Sciences}\ }\textbf {\bibinfo
  {volume} {7}},\ \bibinfo {pages} {189} (\bibinfo {year} {1916})}\BibitemShut
  {NoStop}%
\bibitem [{\citenamefont {Bombelli}\ \emph {et~al.}(1987)\citenamefont
  {Bombelli}, \citenamefont {Lee}, \citenamefont {Meyer},\ and\ \citenamefont
  {Sorkin}}]{Bombelli:1987im}%
  \BibitemOpen
  \bibfield  {author} {\bibinfo {author} {\bibfnamefont {L.}~\bibnamefont
  {Bombelli}}, \bibinfo {author} {\bibfnamefont {J.}~\bibnamefont {Lee}},
  \bibinfo {author} {\bibfnamefont {D.}~\bibnamefont {Meyer}}, \ and\ \bibinfo
  {author} {\bibfnamefont {R.~D.}\ \bibnamefont {Sorkin}},\ }\href {\doibase
  10.1103/PhysRevLett.59.521} {\bibfield  {journal} {\bibinfo  {journal} {Phys.
  Rev. Lett.}\ }\textbf {\bibinfo {volume} {59}},\ \bibinfo {pages} {521}
  (\bibinfo {year} {1987})}\BibitemShut {NoStop}%
\bibitem [{\citenamefont {Bombelli}\ \emph {et~al.}(1988)\citenamefont
  {Bombelli}, \citenamefont {Meyer},\ and\ \citenamefont
  {Sorkin}}]{bombelli1988}%
  \BibitemOpen
  \bibfield  {author} {\bibinfo {author} {\bibfnamefont {L.}~\bibnamefont
  {Bombelli}}, \bibinfo {author} {\bibfnamefont {D.}~\bibnamefont {Meyer}}, \
  and\ \bibinfo {author} {\bibfnamefont {R.~D.}\ \bibnamefont {Sorkin}},\
  }\href@noop {} {\bibfield  {journal} {\bibinfo  {journal} {Physical Review
  Letters}\ }\textbf {\bibinfo {volume} {59}},\ \bibinfo {pages} {521}
  (\bibinfo {year} {1988})}\BibitemShut {NoStop}%
\bibitem [{\citenamefont {Meyer}(1989)}]{meyer1989}%
  \BibitemOpen
  \bibfield  {author} {\bibinfo {author} {\bibfnamefont {D.~A.}\ \bibnamefont
  {Meyer}},\ }\href@noop {} {\bibfield  {journal} {\bibinfo  {journal}
  {Physical Review Letters}\ }\textbf {\bibinfo {volume} {56}},\ \bibinfo
  {pages} {904} (\bibinfo {year} {1989})}\BibitemShut {NoStop}%
\bibitem [{\citenamefont {Sorkin}(1991)}]{sorkin1991}%
  \BibitemOpen
  \bibfield  {author} {\bibinfo {author} {\bibfnamefont {R.~D.}\ \bibnamefont
  {Sorkin}},\ }\href@noop {} {\bibfield  {journal} {\bibinfo  {journal}
  {Journal of Mathematical Physics}\ }\textbf {\bibinfo {volume} {36}},\
  \bibinfo {pages} {2147} (\bibinfo {year} {1991})}\BibitemShut {NoStop}%
\bibitem [{\citenamefont {Sorkin}(1997)}]{sorkin1997}%
  \BibitemOpen
  \bibfield  {author} {\bibinfo {author} {\bibfnamefont {R.~D.}\ \bibnamefont
  {Sorkin}},\ }\href@noop {} {\bibfield  {journal} {\bibinfo  {journal}
  {International Journal of Theoretical Physics}\ }\textbf {\bibinfo {volume}
  {36}},\ \bibinfo {pages} {2759} (\bibinfo {year} {1997})}\BibitemShut
  {NoStop}%
\bibitem [{\citenamefont {Rideout}\ and\ \citenamefont
  {Sorkin}(1999)}]{rideout1999}%
  \BibitemOpen
  \bibfield  {author} {\bibinfo {author} {\bibfnamefont {D.}~\bibnamefont
  {Rideout}}\ and\ \bibinfo {author} {\bibfnamefont {R.~D.}\ \bibnamefont
  {Sorkin}},\ }\href@noop {} {\bibfield  {journal} {\bibinfo  {journal}
  {Physical Review D}\ }\textbf {\bibinfo {volume} {61}},\ \bibinfo {pages}
  {024002} (\bibinfo {year} {1999})}\BibitemShut {NoStop}%
\bibitem [{\citenamefont {Sorkin}(2005{\natexlab{a}})}]{sorkin2005causal}%
  \BibitemOpen
  \bibfield  {author} {\bibinfo {author} {\bibfnamefont {R.~D.}\ \bibnamefont
  {Sorkin}},\ }in\ \href@noop {} {\emph {\bibinfo {booktitle} {Lectures on
  quantum gravity}}}\ (\bibinfo  {publisher} {Springer},\ \bibinfo {year}
  {2005})\ pp.\ \bibinfo {pages} {305--327}\BibitemShut {NoStop}%
\bibitem [{\citenamefont {Sorkin}(2005{\natexlab{b}})}]{sorkin2005}%
  \BibitemOpen
  \bibfield  {author} {\bibinfo {author} {\bibfnamefont {R.~D.}\ \bibnamefont
  {Sorkin}},\ }\href@noop {} {\bibfield  {journal} {\bibinfo  {journal}
  {General Relativity and Gravitation}\ }\textbf {\bibinfo {volume} {38}},\
  \bibinfo {pages} {195} (\bibinfo {year} {2005}{\natexlab{b}})}\BibitemShut
  {NoStop}%
\bibitem [{\citenamefont {Henson}(2006)}]{henson2006}%
  \BibitemOpen
  \bibfield  {author} {\bibinfo {author} {\bibfnamefont {J.}~\bibnamefont
  {Henson}},\ }\href@noop {} {\bibfield  {journal} {\bibinfo  {journal}
  {Foundations of Physics}\ }\textbf {\bibinfo {volume} {36}},\ \bibinfo
  {pages} {545} (\bibinfo {year} {2006})}\BibitemShut {NoStop}%
\bibitem [{\citenamefont {Dowker}\ and\ \citenamefont
  {Surya}(2006)}]{dowker2006}%
  \BibitemOpen
  \bibfield  {author} {\bibinfo {author} {\bibfnamefont {F.}~\bibnamefont
  {Dowker}}\ and\ \bibinfo {author} {\bibfnamefont {S.}~\bibnamefont {Surya}},\
  }\href@noop {} {\bibfield  {journal} {\bibinfo  {journal} {Physics Letters
  A}\ }\textbf {\bibinfo {volume} {357}},\ \bibinfo {pages} {11} (\bibinfo
  {year} {2006})}\BibitemShut {NoStop}%
\bibitem [{\citenamefont {Surya}(2008)}]{surya2008}%
  \BibitemOpen
  \bibfield  {author} {\bibinfo {author} {\bibfnamefont {S.}~\bibnamefont
  {Surya}},\ }\href@noop {} {\bibfield  {journal} {\bibinfo  {journal} {Pramana
  - Journal of Physics}\ }\textbf {\bibinfo {volume} {71}},\ \bibinfo {pages}
  {57} (\bibinfo {year} {2008})}\BibitemShut {NoStop}%
\bibitem [{\citenamefont {Henson}(2010)}]{henson2010}%
  \BibitemOpen
  \bibfield  {author} {\bibinfo {author} {\bibfnamefont {J.}~\bibnamefont
  {Henson}},\ }\href@noop {} {\bibfield  {journal} {\bibinfo  {journal}
  {Entropy}\ }\textbf {\bibinfo {volume} {12}},\ \bibinfo {pages} {1231}
  (\bibinfo {year} {2010})}\BibitemShut {NoStop}%
\bibitem [{\citenamefont {Surya}(2012{\natexlab{a}})}]{surya2012}%
  \BibitemOpen
  \bibfield  {author} {\bibinfo {author} {\bibfnamefont {S.}~\bibnamefont
  {Surya}},\ }\href@noop {} {\bibfield  {journal} {\bibinfo  {journal} {General
  Relativity and Gravitation}\ }\textbf {\bibinfo {volume} {44}},\ \bibinfo
  {pages} {2149} (\bibinfo {year} {2012}{\natexlab{a}})}\BibitemShut {NoStop}%
\bibitem [{\citenamefont {Surya}(2019)}]{Surya:2019aa}%
  \BibitemOpen
  \bibfield  {author} {\bibinfo {author} {\bibfnamefont {S.}~\bibnamefont
  {Surya}},\ }\href {\doibase 10.1007/s41114-019-0023-1} {\bibfield  {journal}
  {\bibinfo  {journal} {Living Reviews in Relativity}\ }\textbf {\bibinfo
  {volume} {22}} (\bibinfo {year} {2019}),\
  10.1007/s41114-019-0023-1}\BibitemShut {NoStop}%
\bibitem [{\citenamefont {Hawking}\ \emph {et~al.}(1976)\citenamefont
  {Hawking}, \citenamefont {King},\ and\ \citenamefont
  {McCarthy}}]{hawking_king_mccarthy_1976}%
  \BibitemOpen
  \bibfield  {author} {\bibinfo {author} {\bibfnamefont {S.~W.}\ \bibnamefont
  {Hawking}}, \bibinfo {author} {\bibfnamefont {A.~R.}\ \bibnamefont {King}}, \
  and\ \bibinfo {author} {\bibfnamefont {P.~J.}\ \bibnamefont {McCarthy}},\
  }\href@noop {} {\bibfield  {journal} {\bibinfo  {journal} {Journal of
  Mathematical Physics}\ }\textbf {\bibinfo {volume} {17}},\ \bibinfo {pages}
  {174} (\bibinfo {year} {1976})}\BibitemShut {NoStop}%
\bibitem [{\citenamefont {Malament}(1977)}]{malament1977}%
  \BibitemOpen
  \bibfield  {author} {\bibinfo {author} {\bibfnamefont {D.}~\bibnamefont
  {Malament}},\ }\href@noop {} {\bibfield  {journal} {\bibinfo  {journal}
  {Journal of Mathematical Physics}\ }\textbf {\bibinfo {volume} {18}},\
  \bibinfo {pages} {1399} (\bibinfo {year} {1977})}\BibitemShut {NoStop}%
\bibitem [{\citenamefont {Brightwell}\ and\ \citenamefont
  {Gregory}(1991)}]{Brightwell:1991aa}%
  \BibitemOpen
  \bibfield  {author} {\bibinfo {author} {\bibfnamefont {G.}~\bibnamefont
  {Brightwell}}\ and\ \bibinfo {author} {\bibfnamefont {R.}~\bibnamefont
  {Gregory}},\ }\href {\doibase 10.1103/PhysRevLett.66.260} {\bibfield
  {journal} {\bibinfo  {journal} {Phys. Rev. Lett.}\ }\textbf {\bibinfo
  {volume} {66}},\ \bibinfo {pages} {260} (\bibinfo {year} {1991})}\BibitemShut
  {NoStop}%
\bibitem [{\citenamefont {Rideout}\ and\ \citenamefont
  {Wallden}(2009)}]{Rideout_2009}%
  \BibitemOpen
  \bibfield  {author} {\bibinfo {author} {\bibfnamefont {D.}~\bibnamefont
  {Rideout}}\ and\ \bibinfo {author} {\bibfnamefont {P.}~\bibnamefont
  {Wallden}},\ }\href {\doibase 10.1088/0264-9381/26/15/155013} {\bibfield
  {journal} {\bibinfo  {journal} {Classical and Quantum Gravity}\ }\textbf
  {\bibinfo {volume} {26}},\ \bibinfo {pages} {155013} (\bibinfo {year}
  {2009})}\BibitemShut {NoStop}%
\bibitem [{\citenamefont {Eichhorn}\ \emph
  {et~al.}(2019{\natexlab{a}})\citenamefont {Eichhorn}, \citenamefont {Surya},\
  and\ \citenamefont {Versteegen}}]{Eichhorn_2019}%
  \BibitemOpen
  \bibfield  {author} {\bibinfo {author} {\bibfnamefont {A.}~\bibnamefont
  {Eichhorn}}, \bibinfo {author} {\bibfnamefont {S.}~\bibnamefont {Surya}}, \
  and\ \bibinfo {author} {\bibfnamefont {F.}~\bibnamefont {Versteegen}},\
  }\href {\doibase 10.1088/1361-6382/ab114b} {\bibfield  {journal} {\bibinfo
  {journal} {Classical and Quantum Gravity}\ }\textbf {\bibinfo {volume}
  {36}},\ \bibinfo {pages} {105005} (\bibinfo {year}
  {2019}{\natexlab{a}})}\BibitemShut {NoStop}%
\bibitem [{\citenamefont {Major}\ \emph {et~al.}(2007)\citenamefont {Major},
  \citenamefont {Rideout},\ and\ \citenamefont {Surya}}]{major2007recovering}%
  \BibitemOpen
  \bibfield  {author} {\bibinfo {author} {\bibfnamefont {S.}~\bibnamefont
  {Major}}, \bibinfo {author} {\bibfnamefont {D.}~\bibnamefont {Rideout}}, \
  and\ \bibinfo {author} {\bibfnamefont {S.}~\bibnamefont {Surya}},\
  }\href@noop {} {\bibfield  {journal} {\bibinfo  {journal} {Journal of
  mathematical physics}\ }\textbf {\bibinfo {volume} {48}} (\bibinfo {year}
  {2007})}\BibitemShut {NoStop}%
\bibitem [{\citenamefont {Brightwell}\ \emph {et~al.}(2008)\citenamefont
  {Brightwell}, \citenamefont {Henson},\ and\ \citenamefont
  {Surya}}]{brightwell20082d}%
  \BibitemOpen
  \bibfield  {author} {\bibinfo {author} {\bibfnamefont {G.}~\bibnamefont
  {Brightwell}}, \bibinfo {author} {\bibfnamefont {J.}~\bibnamefont {Henson}},
  \ and\ \bibinfo {author} {\bibfnamefont {S.}~\bibnamefont {Surya}},\
  }\href@noop {} {\bibfield  {journal} {\bibinfo  {journal} {Classical and
  Quantum Gravity}\ }\textbf {\bibinfo {volume} {25}},\ \bibinfo {pages}
  {105025} (\bibinfo {year} {2008})}\BibitemShut {NoStop}%
\bibitem [{\citenamefont {Major}\ \emph {et~al.}(2009)\citenamefont {Major},
  \citenamefont {Rideout},\ and\ \citenamefont {Surya}}]{Major_2009}%
  \BibitemOpen
  \bibfield  {author} {\bibinfo {author} {\bibfnamefont {S.}~\bibnamefont
  {Major}}, \bibinfo {author} {\bibfnamefont {D.}~\bibnamefont {Rideout}}, \
  and\ \bibinfo {author} {\bibfnamefont {S.}~\bibnamefont {Surya}},\ }\href
  {\doibase 10.1088/0264-9381/26/17/175008} {\bibfield  {journal} {\bibinfo
  {journal} {Classical and Quantum Gravity}\ }\textbf {\bibinfo {volume}
  {26}},\ \bibinfo {pages} {175008} (\bibinfo {year} {2009})}\BibitemShut
  {NoStop}%
\bibitem [{\citenamefont {Benincasa}\ and\ \citenamefont
  {Dowker}(2010)}]{Benincasa:2010eu}%
  \BibitemOpen
  \bibfield  {author} {\bibinfo {author} {\bibfnamefont {D.~M.~T.}\
  \bibnamefont {Benincasa}}\ and\ \bibinfo {author} {\bibfnamefont
  {F.}~\bibnamefont {Dowker}},\ }\href {\doibase
  10.1103/PhysRevLett.104.181301} {\bibfield  {journal} {\bibinfo  {journal}
  {Phys. Rev. Lett.}\ }\textbf {\bibinfo {volume} {104}},\ \bibinfo {pages}
  {181301} (\bibinfo {year} {2010})}\BibitemShut {NoStop}%
\bibitem [{\citenamefont {Surya}(2012{\natexlab{b}})}]{surya2012evidence}%
  \BibitemOpen
  \bibfield  {author} {\bibinfo {author} {\bibfnamefont {S.}~\bibnamefont
  {Surya}},\ }\href@noop {} {\bibfield  {journal} {\bibinfo  {journal}
  {Classical and Quantum Gravity}\ }\textbf {\bibinfo {volume} {29}},\ \bibinfo
  {pages} {132001} (\bibinfo {year} {2012}{\natexlab{b}})}\BibitemShut
  {NoStop}%
\bibitem [{\citenamefont {Saravani}\ and\ \citenamefont
  {Aslanbeigi}(2014)}]{saravani2014causal}%
  \BibitemOpen
  \bibfield  {author} {\bibinfo {author} {\bibfnamefont {M.}~\bibnamefont
  {Saravani}}\ and\ \bibinfo {author} {\bibfnamefont {S.}~\bibnamefont
  {Aslanbeigi}},\ }\href@noop {} {\bibfield  {journal} {\bibinfo  {journal}
  {Classical and Quantum Gravity}\ }\textbf {\bibinfo {volume} {31}},\ \bibinfo
  {pages} {205013} (\bibinfo {year} {2014})}\BibitemShut {NoStop}%
\bibitem [{\citenamefont {Belenchia}\ \emph {et~al.}(2016)\citenamefont
  {Belenchia}, \citenamefont {Benincasa},\ and\ \citenamefont
  {Dowker}}]{belenchia2016continuum}%
  \BibitemOpen
  \bibfield  {author} {\bibinfo {author} {\bibfnamefont {A.}~\bibnamefont
  {Belenchia}}, \bibinfo {author} {\bibfnamefont {D.~M.}\ \bibnamefont
  {Benincasa}}, \ and\ \bibinfo {author} {\bibfnamefont {F.}~\bibnamefont
  {Dowker}},\ }\href@noop {} {\bibfield  {journal} {\bibinfo  {journal}
  {Classical and Quantum Gravity}\ }\textbf {\bibinfo {volume} {33}},\ \bibinfo
  {pages} {245018} (\bibinfo {year} {2016})}\BibitemShut {NoStop}%
\bibitem [{\citenamefont {Machet}\ and\ \citenamefont
  {Wang}(2020)}]{machet2020continuum}%
  \BibitemOpen
  \bibfield  {author} {\bibinfo {author} {\bibfnamefont {L.}~\bibnamefont
  {Machet}}\ and\ \bibinfo {author} {\bibfnamefont {J.}~\bibnamefont {Wang}},\
  }\href@noop {} {\bibfield  {journal} {\bibinfo  {journal} {Classical and
  Quantum Gravity}\ }\textbf {\bibinfo {volume} {38}},\ \bibinfo {pages}
  {015010} (\bibinfo {year} {2020})}\BibitemShut {NoStop}%
\bibitem [{\citenamefont {Bollob{\'a}s}\ and\ \citenamefont
  {Brightwell}(1991)}]{Bollobas:1991aa}%
  \BibitemOpen
  \bibfield  {author} {\bibinfo {author} {\bibfnamefont {B.}~\bibnamefont
  {Bollob{\'a}s}}\ and\ \bibinfo {author} {\bibfnamefont {G.}~\bibnamefont
  {Brightwell}},\ }\href {http://www.jstor.org/stable/2001495} {\bibfield
  {journal} {\bibinfo  {journal} {Transactions of the American Mathematical
  Society}\ }\textbf {\bibinfo {volume} {324}},\ \bibinfo {pages} {59}
  (\bibinfo {year} {1991})}\BibitemShut {NoStop}%
\bibitem [{\citenamefont {Bachmat}(2008)}]{Bachmat:2008aa}%
  \BibitemOpen
  \bibfield  {author} {\bibinfo {author} {\bibfnamefont {E.}~\bibnamefont
  {Bachmat}},\ }\href {\doibase 10.1090/conm/458/08946} {\bibfield  {journal}
  {\bibinfo  {journal} {Contemporary Mathematics}\ }\textbf {\bibinfo {volume}
  {458}} (\bibinfo {year} {2008}),\ 10.1090/conm/458/08946}\BibitemShut
  {NoStop}%
\bibitem [{Note1()}]{Note1}%
  \BibitemOpen
  \bibinfo {note} {Our numerical simulations seems to suggest that in the limit
  $\rho \rightarrow \infty $ $\alpha _1=\alpha _2=\alpha _3$.}\BibitemShut
  {Stop}%
\bibitem [{Note2()}]{Note2}%
  \BibitemOpen
  \bibinfo {note} {To derive this expression, we use error propagation from
  Eq.~\protect \eqref {eq:overlapinc} taking into account that $N_c(a,b)
  \protect \mbox { and } \Delta N_c(a,b)$ are independent Poisson random
  variables.}\BibitemShut {Stop}%
\bibitem [{Note3()}]{Note3}%
  \BibitemOpen
  \bibinfo {note} {Without loss of generality, we assume $a_0 \prec
  a_m$}\BibitemShut {NoStop}%
\bibitem [{\citenamefont {Eichhorn}\ \emph
  {et~al.}(2019{\natexlab{b}})\citenamefont {Eichhorn}, \citenamefont {Surya},\
  and\ \citenamefont {Versteegen}}]{Eichhorn_2019b}%
  \BibitemOpen
  \bibfield  {author} {\bibinfo {author} {\bibfnamefont {A.}~\bibnamefont
  {Eichhorn}}, \bibinfo {author} {\bibfnamefont {S.}~\bibnamefont {Surya}}, \
  and\ \bibinfo {author} {\bibfnamefont {F.}~\bibnamefont {Versteegen}},\
  }\href {\doibase 10.1088/1361-6382/ab47cd} {\bibfield  {journal} {\bibinfo
  {journal} {Classical and Quantum Gravity}\ }\textbf {\bibinfo {volume}
  {36}},\ \bibinfo {pages} {235013} (\bibinfo {year}
  {2019}{\natexlab{b}})}\BibitemShut {NoStop}%
\bibitem [{\citenamefont {Meyer}(1988)}]{Meyer:1988}%
  \BibitemOpen
  \bibfield  {author} {\bibinfo {author} {\bibfnamefont {D.}~\bibnamefont
  {Meyer}},\ }\emph {\bibinfo {title} {The dimension of causal sets}},\
  \href@noop {} {Ph.D. thesis},\ \bibinfo  {school} {M.I.T} (\bibinfo {year}
  {1988})\BibitemShut {NoStop}%
\bibitem [{\citenamefont {Bombelli}(1987)}]{Mid:point}%
  \BibitemOpen
  \bibfield  {author} {\bibinfo {author} {\bibfnamefont {L.}~\bibnamefont
  {Bombelli}},\ }\emph {\bibinfo {title} {Space-time as a causal set}},\
  \href@noop {} {Ph.D. thesis},\ \bibinfo  {school} {Syracuse University}
  (\bibinfo {year} {1987})\BibitemShut {NoStop}%
\bibitem [{\citenamefont {Reid}(2003)}]{Reid:2003kc}%
  \BibitemOpen
  \bibfield  {author} {\bibinfo {author} {\bibfnamefont {D.~D.}\ \bibnamefont
  {Reid}},\ }\href {\doibase 10.1103/PhysRevD.67.024034} {\bibfield  {journal}
  {\bibinfo  {journal} {Phys. Rev. D}\ }\textbf {\bibinfo {volume} {67}},\
  \bibinfo {pages} {024034} (\bibinfo {year} {2003})}\BibitemShut {NoStop}%
\bibitem [{\citenamefont {Belkin}\ and\ \citenamefont
  {Niyogi}(2003)}]{Belkin:2003aa}%
  \BibitemOpen
  \bibfield  {author} {\bibinfo {author} {\bibfnamefont {M.}~\bibnamefont
  {Belkin}}\ and\ \bibinfo {author} {\bibfnamefont {P.}~\bibnamefont
  {Niyogi}},\ }\href {\doibase 10.1162/089976603321780317} {\bibfield
  {journal} {\bibinfo  {journal} {Neural Comput.}\ }\textbf {\bibinfo {volume}
  {15}},\ \bibinfo {pages} {1373} (\bibinfo {year} {2003})}\BibitemShut
  {NoStop}%
\bibitem [{\citenamefont {Glaser}\ and\ \citenamefont
  {Surya}(2013)}]{Glaser:2013hb}%
  \BibitemOpen
  \bibfield  {author} {\bibinfo {author} {\bibfnamefont {L.}~\bibnamefont
  {Glaser}}\ and\ \bibinfo {author} {\bibfnamefont {S.}~\bibnamefont {Surya}},\
  }\href {\doibase 10.1103/PhysRevD.88.124026} {\bibfield  {journal} {\bibinfo
  {journal} {Phys. Rev. D}\ }\textbf {\bibinfo {volume} {88}},\ \bibinfo
  {pages} {124026} (\bibinfo {year} {2013})}\BibitemShut {NoStop}%
\bibitem [{\citenamefont {Wald}(2010)}]{wald2010general}%
  \BibitemOpen
  \bibfield  {author} {\bibinfo {author} {\bibfnamefont {R.}~\bibnamefont
  {Wald}},\ }\href {https://books.google.es/books?id=9S-hzg6-moYC} {\emph
  {\bibinfo {title} {General Relativity}}}\ (\bibinfo  {publisher} {University
  of Chicago Press},\ \bibinfo {year} {2010})\BibitemShut {NoStop}%
\bibitem [{\citenamefont {Varadarajan}\ and\ \citenamefont
  {Rideout}(2006)}]{Varadarajan:2006lq}%
  \BibitemOpen
  \bibfield  {author} {\bibinfo {author} {\bibfnamefont {M.}~\bibnamefont
  {Varadarajan}}\ and\ \bibinfo {author} {\bibfnamefont {D.}~\bibnamefont
  {Rideout}},\ }\href {\doibase 10.1103/PhysRevD.73.104021} {\bibfield
  {journal} {\bibinfo  {journal} {Phys. Rev. D}\ }\textbf {\bibinfo {volume}
  {73}},\ \bibinfo {pages} {104021} (\bibinfo {year} {2006})}\BibitemShut
  {NoStop}%
\bibitem [{\citenamefont {Ahmed}\ and\ \citenamefont
  {Rideout}(2010)}]{Ahmed:2010vn}%
  \BibitemOpen
  \bibfield  {author} {\bibinfo {author} {\bibfnamefont {M.}~\bibnamefont
  {Ahmed}}\ and\ \bibinfo {author} {\bibfnamefont {D.}~\bibnamefont
  {Rideout}},\ }\href {\doibase 10.1103/PhysRevD.81.083528} {\bibfield
  {journal} {\bibinfo  {journal} {Phys. Rev. D}\ }\textbf {\bibinfo {volume}
  {81}},\ \bibinfo {pages} {083528} (\bibinfo {year} {2010})}\BibitemShut
  {NoStop}%
\bibitem [{\citenamefont {Surya}\ and\ \citenamefont
  {Zalel}(2020)}]{Surya_2020}%
  \BibitemOpen
  \bibfield  {author} {\bibinfo {author} {\bibfnamefont {S.}~\bibnamefont
  {Surya}}\ and\ \bibinfo {author} {\bibfnamefont {S.}~\bibnamefont {Zalel}},\
  }\href {\doibase 10.1088/1361-6382/ab987f} {\bibfield  {journal} {\bibinfo
  {journal} {Classical and Quantum Gravity}\ }\textbf {\bibinfo {volume}
  {37}},\ \bibinfo {pages} {195030} (\bibinfo {year} {2020})}\BibitemShut
  {NoStop}%
\bibitem [{\citenamefont {Bento}\ \emph {et~al.}(2022)\citenamefont {Bento},
  \citenamefont {Dowker},\ and\ \citenamefont {Zalel}}]{Bento_2022}%
  \BibitemOpen
  \bibfield  {author} {\bibinfo {author} {\bibfnamefont {B.~V.}\ \bibnamefont
  {Bento}}, \bibinfo {author} {\bibfnamefont {F.}~\bibnamefont {Dowker}}, \
  and\ \bibinfo {author} {\bibfnamefont {S.}~\bibnamefont {Zalel}},\ }\href
  {\doibase 10.1088/1361-6382/ac445f} {\bibfield  {journal} {\bibinfo
  {journal} {Classical and Quantum Gravity}\ }\textbf {\bibinfo {volume}
  {39}},\ \bibinfo {pages} {045002} (\bibinfo {year} {2022})}\BibitemShut
  {NoStop}%
\bibitem [{\citenamefont {Zalel}(2023)}]{Zalel2023}%
  \BibitemOpen
  \bibfield  {author} {\bibinfo {author} {\bibfnamefont {S.}~\bibnamefont
  {Zalel}},\ }\enquote {\bibinfo {title} {Covariant growth dynamics},}\ in\
  \href {\doibase 10.1007/978-981-19-3079-9_82-1} {\emph {\bibinfo {booktitle}
  {Handbook of Quantum Gravity}}},\ \bibinfo {editor} {edited by\ \bibinfo
  {editor} {\bibfnamefont {C.}~\bibnamefont {Bambi}}, \bibinfo {editor}
  {\bibfnamefont {L.}~\bibnamefont {Modesto}}, \ and\ \bibinfo {editor}
  {\bibfnamefont {I.}~\bibnamefont {Shapiro}}}\ (\bibinfo  {publisher}
  {Springer Nature Singapore},\ \bibinfo {address} {Singapore},\ \bibinfo
  {year} {2023})\ pp.\ \bibinfo {pages} {1--36}\BibitemShut {NoStop}%
\bibitem [{\citenamefont {Last}\ and\ \citenamefont
  {Penrose}(2017)}]{last2017lectures}%
  \BibitemOpen
  \bibfield  {author} {\bibinfo {author} {\bibfnamefont {G.}~\bibnamefont
  {Last}}\ and\ \bibinfo {author} {\bibfnamefont {M.}~\bibnamefont {Penrose}},\
  }\href {\doibase 10.1017/9781316104477} {\emph {\bibinfo {title} {{Lectures
  on the Poisson Process}}}}\ (\bibinfo  {publisher} {Cambridge University
  Press},\ \bibinfo {address} {Cambridge, UK},\ \bibinfo {year}
  {2017})\BibitemShut {NoStop}%
\end{thebibliography}

%

\end{document}